%% file: ME895rv.tex
\documentclass[usegraphicx,useAMS]{mn2e}
\usepackage{subfigure}

\title[Low mass stars and brown dwarfs in the $\sigma$~Ori cluster]
  {Membership, binarity and accretion
  among very low-mass stars
  and brown dwarfs of the $\sigma$~Orionis cluster}
  
\author[M.J.~Kenyon, et al.]
  {M.J.~Kenyon$^1$, R.D.~Jeffries$^1$, Tim Naylor$^2$, J.M.~Oliveira$^1$
  and P.F.L.~Maxted$^{1}$\\
  $^1$School of Chemistry and Physics, Keele University, Keele, 
      Staffordshire ST5 5BG, United Kingdom\\
$^2$School of Physics, Stocker Road, University of Exeter, Exeter EX4 4QL, United Kingdom\\
}
\date{Released September 30 2004}

\pagerange{\pageref{firstpage}--\pageref{lastpage}} \pubyear{2004}

\def\LaTeX{L\kern-.36em\raise.3ex\hbox{a}\kern-.15em
    T\kern-.1667em\lower.7ex\hbox{E}\kern-.125emX}

\newcommand{\lith}{Li\,{\sc i}\, }
\newcommand{\sod}{Na\,{\sc i}\, }

\begin{document}

\label{firstpage}

\maketitle

\begin{abstract}
Intermediate resolution ($R\sim 7000$) spectroscopy is presented for 76
photometrically selected very low mass ($0.04<M<0.3\,M_{\odot}$)
candidate members of the young cluster around $\sigma$~Orionis. More
than two thirds appear to be genuine cluster members on the basis that they
exhibit \lith 6708\AA\ absorption, weak \sod
8183/8195\AA\ features and a radial velocity consistent with the
cluster mean. Photometric selection alone therefore appears to be very
effective in identifying cluster members in this mass range.
Only 6 objects appear to be certain non-members, however a substantial 
subset of 13 have ambiguous or contradictory
indications of membership and lack Li absorption. This together with an
observed spread in the equivalent width of the Li absorption feature in
the cooler stars of our sample indicates there may be deficiencies in
our understanding of the formation of this line in cool, low-gravity
objects.
 
Four candidate binary cluster members are identified. Consideration of
sampling and radial velocity measurement precision leads us to conclude
that either the fraction of very low mass stars and brown dwarfs in
small separation ($a<1$\,au) binary systems is larger than in field
M-dwarfs, or the distribution of separations is much less skewed
towards large separations. This conclusion hinges critically on the
correct identification of the small number of binary candidates,
although it remains significant even when only the candidate members
displaying Li absorption are considered. 

Broadened H$\alpha$ emission, indicative of circum(sub)stellar
accretion discs is found in 5 or 6 of the candidate cluster members, 3
of which probably have substellar masses. The fraction of accretors
($10\pm5$ per cent) is similar to that found in stars of higher mass in
the $\sigma$~Ori cluster using H$\alpha$ emission as a diagnostic, but
much lower than found for very low mass stars and brown dwarfs of
younger clusters. The timescale for accretion rates to drop to $\la
10^{-11}\,M_{\odot}$\,yr$^{-1}$ is hence less than the age of the
$\sigma$~Ori cluster (3 to 7\,Myr) for most low-mass objects.

\end{abstract}


\section{Introduction}
The typical Jeans mass
in molecular cloud cores is an order of magnitude more massive than a
brown dwarf, yet somehow these substellar runts are found in great
abundance -- either isolated in the field, as companions to more massive
objects or as members of young clusters (e.g. Bouvier et al. 1998;
Kirkpatrick et al. 2000; B\'ejar et al. 2001, Gizis et al. 2001; Moraux
et al. 2001).  Despite this, there is no
universally accepted formation scenario which explains the profusion of
substellar objects. Some theories propose that turbulent
fragmentation can extend the more conventional star formation scenario
to low masses (Padoan \& Nordlund 2002), whilst others have proposed
that BDs can form through instabilities in circumstellar or
circumbinary discs (Pickett et al. 2000; Jiang, Laughlin \& Lin 2004);
by early ejection from protostellar aggregates (Reipurth \& Clarke
2001); or by removal of accretion envelopes by collision or
photoevaporation (Price \& Podsialowski 1995; Kroupa \& Bouvier 2003).
Possible diagnostics that may point to which processes are the most
important include the initial mass function (IMF) and the properties of
binary systems and discs among the lowest mass objects.

The IMF
is often represented by the power-law relationship;
$dn/d \log m \propto m^{-\alpha}$.  Since the classic work of
Salpeter (1955) the IMF has been well studied down to masses of order
$\sim1$M$_{\sun}$ and is best described by an exponent of $\alpha
=1.35$. Below this it flattens, with $\alpha\simeq 0$ having been
measured in the field and in several young open clusters (see Chabrier
2003 for a review). For VLMS and BDs the IMF is more
uncertain. Most determinations in young clusters report $-0.5<\alpha<0$ 
(e.g. Moraux et al. 2001; B\'ejar et al. 2001), which are consistent with 
determinations in the field once appropriate corrections for unresolved 
binarity are applied (Chabrier 2003). Recently there have been claims for 
non-universality of the IMF at the lowest masses. 
Brice\~no et al. (2002) and Luhman et al. (2003)
report a significantly smaller number of BDs per star
in Taurus-Auriga and IC~348 than in the Orion Nebula cluster, 
perhaps indicating an
important environmental dependence for the BD formation process. 

Determination of the true IMF is inextricably linked with a knowledge of
binary properties. VLMS and BDs can be hidden in unresolved binary
systems, changing the observed mass functions and possibly introducing
apparent IMF variations between clusters with different binary
frequencies (Kroupa \& Bouvier 2003). The statistics of binary systems
themselves -- mass ratios ($q$) and separations ($a$) -- are also important.  
The binary properties of BDs and VLMS are not simply an extension
of those for low-mass stars. The lack of wide binary VLMS and BD systems
($a>20$\,au) in the field and young clusters is notable (Mart{\' \i}n et
al 2003; Bouy et al. 2003; Close et al. 2003; Gizis et al. 2003) and has
been taken as support for dynamical ejection models. The overall binary
frequency is still uncertain. Directly resolved binary systems with
$1.6<a<16$\,au and $q\geq0.5$ form about 15 per cent of the
population. This figure is reasonably consistent with, although a little
higher than, the number of BD binaries with separations $>1$\,au 
formed in hydrodynamical simulations (Bate, Bonnell \& Bromm
2003). Closer binaries are much less well studied. Examples of
spectroscopic BD binaries exist (e.g. PPL 15 -- Basri \& Mart{\' \i}n
1999) and may be quite common (Reid et al. 2002; Guenther \& Wuchterl
2003). An
indirect estimate of the fraction of unresolved binaries in the
Pleiades, using position in colour-magnitude diagrams, led Pinfield et
al (2003) to conclude that the BD binary fraction with $q>0.4$ in the 
Pleiades was $50^{+11}_{-10}$ per cent. The implication is that most of 
these must be close binaries ($a<1.6$\,au) in order to agree with the 
statistics for wider binary systems found in high spatial resolution 
imaging.

Insights into VLMS and BD formation may also come from the presence
(or otherwise) of circumstellar discs, which can be 
characterised by strong H$\alpha$ emission or
infrared excesses. Bate et al. (2003) argue that the ejection of a brown
dwarf will result in the truncation of any surrounding material leading
to less massive discs with shorter lifetimes. However, recent results using
accretion diagnostics and $L^{\prime}$-band excesses suggest that many BDs
 do indeed possess (accretion) discs and that their lifetimes may
be similar to discs around more massive objects (see Liu, Najita \&
Tokunaga 2003; Jayawardhana, Mohanty \& Basri 2003; Jayawardhana et al. 2003b).

To a greater or lesser extent, studies of the IMF, binarity and discs in
young clusters are hampered by a lack of firm membership information,
and in the case of close binarity in VLMS/BDs, a lack of data.
Samples must be free of contamination to draw accurate conclusions
about the IMF and disc frequencies.
Here we explore these issues using the VLMS/BD population clustered
around the O9.5V star $\sigma$~Orionis, which is itself a member of the
Orion OB1b association (Walter et al. 1998).  At an
Hipparcos distance of only $352^{+166}_{-85}$\,pc, aged $3-7$\,Myr
(Oliveira et al. 2002) and with low reddening ($E(B-V)=0.05$\,mag, Brown
et al. 1994), the $\sigma$ Ori cluster is a favourable environment in which to
study VLMS/BD evolution. Its youth guarantees that low-mass members
will be relatively luminous. The VLMS and BD population of $\sigma$ Ori
has been investigated by a number of authors.  B\'ejar et al. (1999,
2001) have combined optical and near-infrared photometric surveys of
the $\sigma$ Ori region to locate candidate VLMS and BDs and estimate
an IMF with $\alpha=-0.2\pm0.4$. There is some evidence that the IMF
extends down to a few Jupiter masses (e.g. Zapatero-Osorio et
al. 2000).  Low resolution spectroscopy, with resolving power
$250<R<500$, has been obtained
for many of these objects (B\'ejar et al. 1999; Zapatero-Osorio et
al. 2000; Barrado y Navascu\'{e}s et al. 2003). Almost all are claimed to be
members based on their spectral types, H$\alpha$ emission
characteristics and for some objects, on the basis that their neutral
alkali lines (K\,{\sc i}, Na\,{\sc i}) are weaker than expected for
dwarf stars, indicating a low gravity. 
In our opinion, only the latter can reasonably discriminate
against contaminating objects in photometrically selected samples. A
spectral type consistent with colour {\em is not} evidence for
membership because the dominant contaminants will be field M dwarfs
with similar spectral types. Nor is moderate H$\alpha$ emission
conclusive. A large fraction of field M-dwarfs show H$\alpha$ emission,
especially in the range M3-M6V (Gizis, Reid \& Hawley 2002).

Higher resolution spectroscopy offers a number of advantages in
determining membership credentials. The presence of Li in the
atmosphere can be diagnosed from the \lith 6708\AA\ line, indicating a
youthful star. This technique works because Li is burned very quickly
in fully convective VLMS envelopes once the core reaches
$\sim3\times10^{6}$\,K. Furthermore, radial velocities (RVs) allow the 
kinematic properties of candidates to be used as a check on membership. 
Little work at high resolution has been done on VLMS and BDs around 
$\sigma$ Ori. Zapatero-Osorio et al. (2002) obtained spectroscopy of 
26 low-mass stars and 2 BD candidates at $1600<R<4400$, of which only 5 
had $I>15$, corresponding to masses $<0.2M_{\odot}$. These data
demonstrated that Li should remained unburned in {\em all} VLMS and BD 
cluster members. Muzerolle et al. (2003) obtained $R=8000$ and $R=34000$
spectroscopy of 7 BD candidates from B\'ejar et al. (1999, 2001) 
with $16.5<I<19.8$, finding that 2 have radial velocities inconsistent
with cluster membership and showing no Li. Finally, McGovern et al. (2004) 
observed 2 photometric BD candidates at $R=1000-2000$ in the near infrared, 
showing that one object had neutral alkali lines too strong to be a 
pre-main-sequence (PMS) object.

In this paper we present new $RI$ photometry and $R\simeq7000$ optical
and near IR spectroscopy of 76 photometrically selected VLMS/BD
candidates around $\sigma$ Ori with $14.8<I<18.2$. Our objectives are
as follows.
\begin{enumerate}
\item To confirm
membership of the association for these objects, 
using radial velocities and the strength of Li\,{\sc i} and
Na\,{\sc i} lines, to validate IMFs derived solely from
photometrically selected members (e.g. B\'ejar et al. 2001). 
\item To search for short period binary systems to see whether 
there are a large number of VLMS and BDs hidden in such systems as 
suggested by Pinfield et al. (2003).
\item To look for the presence of accretion betrayed by the
  presence of broad H$\alpha$ emission lines and compare disc
  frequencies in the VLMS/BDs of $\sigma$ Ori with those in stars of
  higher mass and with objects of similar mass in younger clusters.
\end{enumerate}

Section 2 of the paper outlines the optical dataset used to select
candidate cluster members; section~3 describes the multi-object 
fibre spectroscopy; section 4 presents the spectroscopic results and
radial velocities; cluster membership is discussed in section 5 and in
section 6 we discuss the use of various membership indicators in star
forming regions and the IMF, binarity and accretion statistics of the
cluster around $\sigma$ Ori. Our conclusions can be found in section~7.

\section{Optical Photometry}
We observed the field around $\sigma$ Ori with the Wide Field Camera
(WFC) on the 2.5-m Isaac Newton Telescope using Harris $R$ and Sloan
$i$ filters on the nights of $27-30$ September 1999.  The WFC consists
of a mosaic of 4 2048$\times$4096 pixel CCDs covering a sky area of
about 970 square arcminutes and we obtained 8 overlapping pointings,
only 5 of which (pointings 1, 2, 3, 4 and 8) are relevant to this
paper. The details of our observations are given in Table~\ref{WFClog}.

\begin{table*}

 \caption{Wide Field Camera observing log. Coordinates refer to the
 rotator centre of the telescope.}
 \label{WFClog}
 \begin{tabular}{@{}lcccccccccc}
\hline
 \multicolumn{1}{l}{}&
 \multicolumn{2}{c}{Pointing 1} &
 \multicolumn{2}{c}{Pointing 2} &
 \multicolumn{2}{c}{Pointing 3} &
 \multicolumn{2}{c}{Pointing 4} &
 \multicolumn{2}{c}{Pointing 8} \\

\hline

 \multicolumn{1}{l}{RA(J2000)}&
 \multicolumn{2}{c}{05 39 50}&
 \multicolumn{2}{c}{05 39 50}&
 \multicolumn{2}{c}{05 37 42}&
 \multicolumn{2}{c}{05 37 44}&
 \multicolumn{2}{c}{05 41 20}\\

 \multicolumn{1}{l}{Dec(J2000)}&
 \multicolumn{2}{c}{-02 20 00}&
 \multicolumn{2}{c}{-02 51 40}&
 \multicolumn{2}{c}{-02 20 00}&
 \multicolumn{2}{c}{-02 51 40}&
 \multicolumn{2}{c}{-02 27 50}\\

\hline

 \multicolumn{1}{l}{Filter}&
 \multicolumn{1}{c}{R}&
 \multicolumn{1}{c}{i}&
 \multicolumn{1}{c}{R}&
 \multicolumn{1}{c}{i}&
 \multicolumn{1}{c}{R}&
 \multicolumn{1}{c}{i}&
 \multicolumn{1}{c}{R}&
 \multicolumn{1}{c}{i}&
 \multicolumn{1}{c}{R}&
 \multicolumn{1}{c}{i} \\

\hline


 \multicolumn{1}{l}{1999-09-27}&
 \multicolumn{1}{c}{2x300s}&
 \multicolumn{1}{c}{2x150s}&
 \multicolumn{1}{c}{2x300s}&
 \multicolumn{1}{c}{2x150s}&
 \multicolumn{1}{c}{}&
 \multicolumn{1}{c}{}&
 \multicolumn{1}{c}{}&
 \multicolumn{1}{c}{}&
 \multicolumn{1}{c}{}&
 \multicolumn{1}{c}{}\\

 \multicolumn{1}{l}{1999-09-28}&
 \multicolumn{1}{c}{1x20s}&
 \multicolumn{1}{c}{1x10s}&
 \multicolumn{1}{c}{1x20s}&
 \multicolumn{1}{c}{1x10s}&
 \multicolumn{1}{c}{1x20s}&
 \multicolumn{1}{c}{1x10s}&
 \multicolumn{1}{c}{1x20s}&
 \multicolumn{1}{c}{1x10s}&
 \multicolumn{1}{c}{}&
 \multicolumn{1}{c}{}\\

\multicolumn{1}{l}{}&
 \multicolumn{1}{c}{}&
 \multicolumn{1}{c}{}&
 \multicolumn{1}{c}{}&
 \multicolumn{1}{c}{}&
 \multicolumn{1}{c}{1x300s}&
 \multicolumn{1}{c}{2x150s}&
 \multicolumn{1}{c}{1x300s}&
 \multicolumn{1}{c}{2x150s}&
 \multicolumn{1}{c}{}&
 \multicolumn{1}{c}{}\\

 \multicolumn{1}{l}{1999-09-29}&
 \multicolumn{1}{c}{}&
 \multicolumn{1}{c}{}&
 \multicolumn{1}{c}{}&
 \multicolumn{1}{c}{}&
 \multicolumn{1}{c}{}&
 \multicolumn{1}{c}{}&
 \multicolumn{1}{c}{1x300s}&
 \multicolumn{1}{c}{1x150s}&
 \multicolumn{1}{c}{}&
 \multicolumn{1}{c}{}\\

 \multicolumn{1}{l}{1999-10-01}&
 \multicolumn{1}{c}{}&
 \multicolumn{1}{c}{}&
 \multicolumn{1}{c}{}&
 \multicolumn{1}{c}{}&
 \multicolumn{1}{c}{}&
 \multicolumn{1}{c}{}&
 \multicolumn{1}{c}{}&
 \multicolumn{1}{c}{}&
 \multicolumn{1}{c}{1x20s}&
 \multicolumn{1}{c}{1x10s}\\

 \multicolumn{1}{l}{}&
 \multicolumn{1}{c}{}&
 \multicolumn{1}{c}{}&
 \multicolumn{1}{c}{}&
 \multicolumn{1}{c}{}&
 \multicolumn{1}{c}{}&
 \multicolumn{1}{c}{}&
 \multicolumn{1}{c}{}&
 \multicolumn{1}{c}{}&
 \multicolumn{1}{c}{3x300s}&
 \multicolumn{1}{c}{3x150s}\\

 \hline
 \end{tabular}
\end{table*}

Each flatfield, bias and data frame was bias subtracted using the mean of an 
over-scan region, and then linearised using the September/October 1999 
coefficients. After this, a median bias frame was subtracted from the 
flats and data frames to remove any residual bias pattern. Since the camera 
was suffering from light leaks during the run, the twilight flats could only 
be used to remove small scale pixel-to-pixel variability. The large scale 
flatfielding was achieved using flats taken in March 1999. Finally all the 
$i$ band data were defringed using master fringe frames constructed from 
data taken in February 1999. These frames were created by normalising and 
then median stacking data frames from each night, and then creating an 
average from several nights. Varying fractions of the fringe frame were then 
subtracted from the data until the mean average deviation in the sky of the 
data frame was minimised. From this point the data reduction followed the 
procedure described in detail in Naylor (1998) and Naylor et al. (2002). 
We created an initial catalogue of objects by searching the sum of the long 
$i$-band exposures, and then carried out optimal photometry in each separate 
CCD image at those positions.

\begin{figure}
\includegraphics[width=84mm]{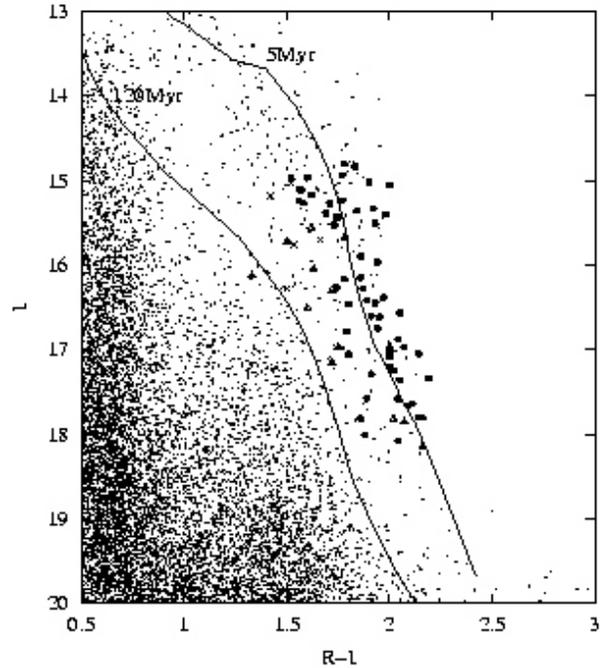}
\caption{I,R-I colour-magnitude diagram. Spectroscopic targets are
 shown as follows: circles represent objects which display lithium in
 their optical spectra, crosses are definite cluster non-members, and
 our ``grey area'' objects are indicated by open triangles (see section
 6.1).  Two isochrones are shown, aged $5$ and $120$Myr, derived using
 the models of Baraffe et al. (1998) and assuming a distance of 352\,pc
 and $E(R-I)=0.03$.  }
\label{ricmd}
\end{figure}

Our astrometry was initially performed using a nine parameter model, which
included the tangent point and pincushion distortion co-efficient as
free parameters. Although we finally used the SuperCosmos sky survey (Hambly et
al. 2001) as our reference catalogue, the stars were so sparse in the
parts of pointing 8 (see Table~1) 
in the direction of the molecular cloud, that for
this pointing only, our initial solutions were with respect to the
2MASS survey (Cutri et al. 2003). We found the mean of the distortion 
co-efficients derived from each CCD image was close to the nominal value, 
so we fixed its value at the nominal 220. We then fitted again to find the 
tangent points for each CCD image, which we then averaged to find a tangent 
point for each pointing. With these parameters fixed, final astrometric 
solutions with respect to SuperCosmos (with just six free parameters) were 
found. The RMS deviation from the astrometric solutions  for each CCD ranged 
between 0.13 and 0.20 arcsec.

To correct our optimal fluxes to those that would have been obtained 
from a large aperture we used second and third order polynomials to
model the profile-correction in the 
directions of the short and long axes of the CCDs respectively (see
Naylor et al. 2002).
The photometric co-efficients used to convert the profile corrected
magnitudes to apparent magnitudes were derived from observations of
over 700 standard stars in $R$ and $i$ during the 4 photometric nights of 
the run (Landolt 1992). These were used to obtain extinction co-efficients 
(1 per night), colour co-efficients (1 per CCD) and zero points (1 per CCD 
per night). However, the small range of airmasses we observed at (1.1 to 1.5) 
meant that we fixed the extinction at the La Palma mean values (0.04 in $I$, 
0.05 in $R-I$). We found we had to add a large systematic error (0.04 mag in $I$
0.06 mag in $R-I$) to obtain a reduced $\chi^{2}$ of 1 for the fit. After doing 
this, our final number of standard stars used was 653 (8 clipped out) in 
$R-I$ and 720 (5 clipped out) in $I$. Note that although we used a Sloan 
$i$ filter, the magnitudes we derived are tied to the Landolt standards and
so they are on the Cousins system.

The resulting apparent magnitudes could then be compared for stars which 
fell in more than one CCD field. From the stars in each overlap region we 
calculated a mean difference for that overlap. We then adjusted the zero 
points to minimise the differences, obtaining a root-mean-square (RMS) 
discrepancy of 0.007 
and 0.008 mags in $I$ and $R-I$ respectively. Where a CCD had no overlaps, 
its photometry was adjusted 
by the mean correction for that pointing, to maintain consistency. Finally,
the mean magnitude was calculated for stars appearing in two or more CCD 
fields.

As explained in Naylor et al. (2002), the overlaps are a powerful diagnostic 
of the accuracy of the photometry, since they test errors in profile 
correction and large-scale flatfielding, in addition to any change in 
transparency. They represent a very good measure of the likely photometric 
error. In this case, before we adjusted the zero points, we found a RMS 
difference of 0.05 mags between the overlaps. Since this closely matches the 
systematic error we had to add to the standard stars, we believe this 
represents our photometric accuracy. We regard this as rather poor, and are 
unsure of why this is so. Some of the error is due to differences between our 
filters and those used to create the standard star system. Specifically, we 
have used a Sloan $i$ filter, whilst the Landolt standards are for $I$.
However, this is unlikely to be the entire cause of the problem, since a 
linear fit of $R-I$ vs $R-i$ for the stars of Smith et al. (2002) gives an
RMS of 0.026 mags. In other work, using the same software, we have found much 
smaller differences between the magnitudes of stars measured in different CCD
pointings (Naylor et al. 2002; Burningham et al. 2003).  This suggests to us 
that the remaining problem originates in the flatfield, although we cannot 
prove that it is not the profile correction which is at fault.
Our defringing technique did not entirely remove the fringes, leaving
structure on a similar scale to our sky background boxes.  This means the
sky subtraction is not perfect, which increases the uncertainties in our
photometry beyond those given by the optimal extraction.  Experiments with
photometry of empty regions of sky showed that the uncertainties we
quote should be increased by a factor of about 1.5.

We compared our catalogue with that of B\'ejar et al. (2001). After applying 
a shift of 1.3 arcsec in RA and 1.1 arcsec in declination to our catalogue, 
we found good correlations out to 5 arcsec, with an RMS separation of 1.8 
arcsec between the two catalogues. Our small astrometric RMS with respect to 
SuperCosmos suggests these numbers represent systematic shifts and 
uncertainties in B\'ejar et al.'s astrometry. Choosing just the objects with 
a signal-to-noise ratio greater than 10 in both catalogues, we found 
$I-I_{Bejar}=-0.12\pm0.04$ and $(R-I)-(R-I)_{Bejar}=-0.07\pm0.04$.
There were no significant colour terms, but the range of colours used is 
small (just the PMS). Of more concern is the scatter in the relationships, 
with the RMS difference between measurement's of the same star in the two 
catalogues being 0.24 magnitudes in $I$ and 0.13 magnitudes in $R-I$.
Of course some of this scatter is likely due to variability as most of the
comparison stars are probably PMS objects that are 
young, rapidly rotating and magnetically active, possibly with
accretion discs (see Scholz \& Eisl\"offel 2004).

\section{Spectroscopy}

\begin{figure*}
\includegraphics[width=125mm, angle=270]{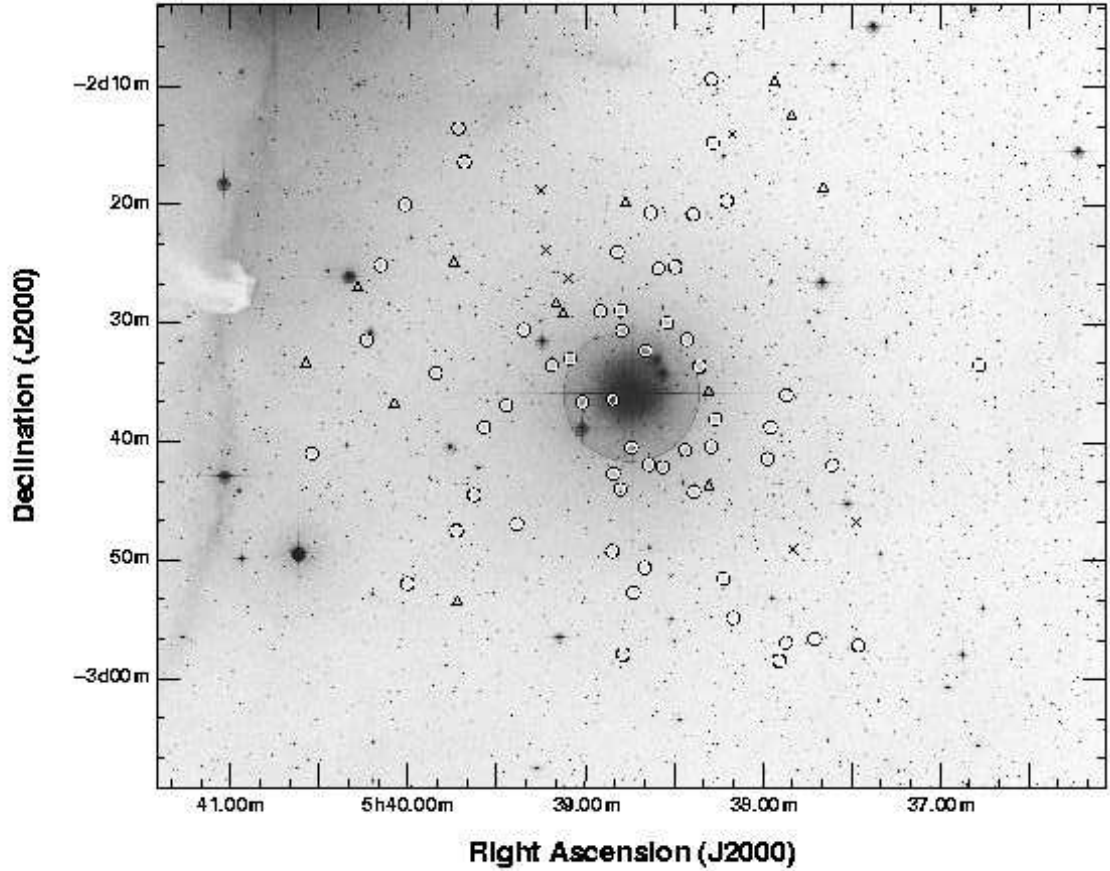}
\caption{Digital Sky Survey image of $\sigma$ Orionis. The positions of spectroscopic 
targets are marked. The symbols have the meanings described in the
caption to Figure 1.
}
\label{imageplot}
\end{figure*}

\subsection{Target selection}

Our spectroscopic targets have been selected on the basis of their
position on the $I,R-I$ colour-magnitude diagram (CMD), see
Figure~\ref{ricmd}.  We chose objects that were reasonably close to a
5\,Myr isochrone, generated from Baraffe et al. (1998) models, using an
empirical colour-effective temperature relationship derived from
Pleiades data and an assumed distance of 352\,pc (see Jeffries, Thurston
\& Hambly 2001 for details of the technique).  The corresponding
objects are highlighted on a Digitized Sky Survey (DSS) image in
Figure~\ref{imageplot}. Targets with $14.8<I<18.2$ (corresponding
approximately to masses $0.04<M<0.3\,M_{\odot}$ using a fixed age of
5\,Myr, a distance of 352\,pc and the Baraffe et al. 1998 models) were
chosen from the catalogue with the aim of maximising the number of
targets close to the cluster isochrone that could be observed in a pair
of multi-fibre setups (see below). Note that because of the limited
number of spectrograph fibres and other geometric restrictions 
(see below) we were unable to place
fibres on all the good cluster candidates in the field of view (see
also section 6.3).

Unfortunately IR photometry was unavailable at the time at which the
spectroscopy was performed, but was later obtained for all our targets 
from the 2MASS all-sky point source catalogue 
(Cutri et al. 2003). 

\subsection{Spectroscopic Observations}
Spectra were obtained with the Wide Field Fiber Optic Spectrograph (WYFFOS)
mounted at the Nasmyth focus of the 4.2-m William Herschel Telescope 
during the nights of 11 and 12 December 1999.
An echelle grating and order sorting filters were 
used to obtain spectra for targets in a
field of diameter just less than 1 degree (see Fig.~\ref{imageplot}), 
covering wavelength ranges of $6390-6810$\AA\ and $7820-8460$\AA\
and at dispersions of 0.43\AA\,pixel$^{-1}$ and
0.63\AA\,pixel$^{-1}$ respectively.
Measurements of arc-line spectra showed that FWHM resolutions of $\sim 0.9$ and $1.3$
\AA\, were attained with these two settings. 
WYFFOS had about 100 active fibres which could be allocated, but
restrictions on the fibre placement and fibre proximity meant that not
all of these could be used. Different fibre/target configurations
were used on the two nights, with 19 (predominantly faint) objects common to 
both configurations. The optical fibres had a diameter of 2.7\,arcsec and to 
estimate the co-temporal sky spectrum we allocated 15 fibres in each 
configuration to blank sky regions.
 
A log of the spectroscopic observations is given in Table~\ref{observ}. In 
addition to several hours of target exposures on each night we obtained 
observations of a tungsten lamp to aid in fibre location and flat fielding, 
Copper plus Neon arc lamp spectra for wavelength calibration at the end of 
each target exposure, several offset exposures of blank sky that were median 
stacked to calibrate the fibre-to-fibre transmission efficiencies in each 
configuration and wavelength setting, spectra of a number of M dwarfs that 
can be used as spectral-type and radial velocity standards and a spectrum 
of a bright, B3V star (HR 5191) that was used to identify and correct for 
telluric absorption lines.

\begin{figure*}
    \centering
    \begin{minipage}{100cm}
    \subfigure[Sample of optical spectra displaying the Li\,{\sc i} line at
    $6708$\AA. Spectra are 
    normalised within the region $6700-6715$\AA\ prior to a constant offset 
    being applied, labels are as given in Table~\ref{rvmembers}. The
    gaps in the spectra correspond to a poorly subtracted bright S\,{\sc
    ii} sky line at 6717\AA.
    ]
    {
    \label{li.montage}\includegraphics[width=84mm]{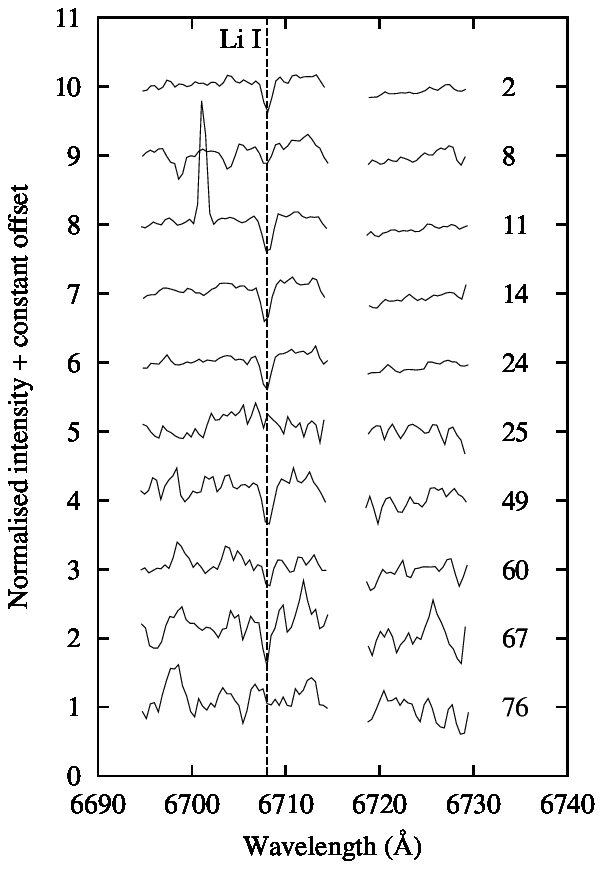}
    }
    \subfigure[Near-infrared spectra showing the 8183,8195\AA\,
    Na\,{\sc i} doublet.  The dashed spectra are of the M6V field dwarf GL412b to
    demonstrate the difference in equivalent widths. Spectra are normalised 
    using the region 8160-8220\AA\ and a constant offset
    applied, labels are as given in Table~\ref{rvmembers}. 
    ]
    {
    \label{na.montage}\includegraphics[width=84mm]{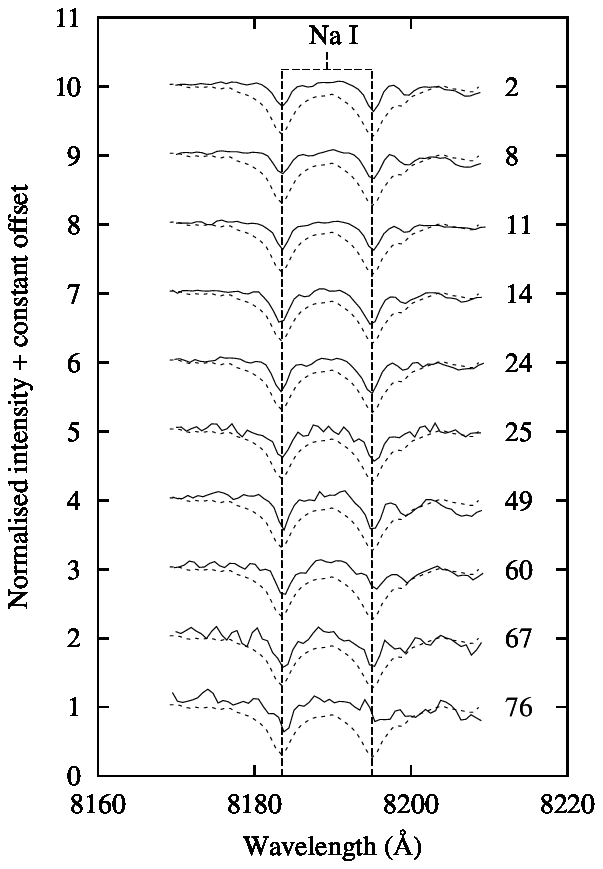}
    }
    \end{minipage}
    \caption{Sample optical and near-infrared spectra.}
\label{montages}
\end{figure*}

\begin{table*}
 \caption{WYFFOS observing Log. Different fibre configurations were employed 
on the two nights, indicated by the subscripts.}
 \label{observ}
 \begin{tabular}{@{}lccccc}
 \hline
          &Target          & RA        & DEC      &Central wavelength&Exposure\\
          &                & (J2000)   & (J2000)  &      (\AA)       &  (s)   \\
 \hline
1999-12-11&$\sigma$-Ori$_1$&05 38 43.44&-02 36 05.0   &6646              &3600\\
          &$\sigma$-Ori$_1$&     ''      &   ''       &6646              &3600\\
          &$\sigma$-Ori$_1$&     ''      &   ''       &6646              &3600\\
          &$\sigma$-Ori$_1$&     ''      &   ''       &6646              &4500\\
          &$\sigma$-Ori$_1$&     ''      &   ''       &8202              &1800\\
          &$\sigma$-Ori$_1$&     ''      &   ''       &8202              &1800\\
1999-12-12&$\sigma$-Ori$_2$&     ''      &   ''       &6649              &4500\\
          &$\sigma$-Ori$_2$&     ''      &   ''       &6649              &4500\\
          &$\sigma$-Ori$_2$&     ''      &   ''       &6649              &3600\\
          &$\sigma$-Ori$_2$&     ''      &   ''       &6649              &3000\\
          &$\sigma$-Ori$_2$&     ''      &   ''       &8200              &2400\\
          &$\sigma$-Ori$_2$&     ''      &   ''       &8205              &1800\\
          &$\sigma$-Ori$_2$&     ''      &   ''       &8205              &1800\\
          &GL 273          &07 24 37.89&05 20 12.2&8199              &20\\
          &GL 388          &10 19 28.69&19 52 36.9&8199              &20\\
          &GL 406          &10 56 21.54&07 01 16.4&8199              &240\\
          &GL 285          &07 44 33.06&03 33 40.5&8199              &40\\
          &GL 412b         &11 05 21.28&43 31 35.2&8199              &120\\
          &GL 447          &11 47 37.03&00 48 37.6&8199              &40\\
          &GL 1156         &12 18 52.02&11 07 53.4&8199              &120\\
          &HR 5191         &13 47 20.37&49 19 14.6&8199              &4\\
 \hline
 \end{tabular}
\end{table*}

\subsection{Data Reduction}

Images were processed using several packages from within the 
{\sc iraf}\footnote{{\sc iraf} is distributed by National Optical Astronomy 
Observatory, which is operated by the Association of Universities for Research 
in Astronomy, Inc., under contract with the National Science Foundation.} 
environment. Most notable of these was {\sc wyf\_red} (Lewis 1996), a script based on the
fibre reduction package {\sc dofibers}, but with some differences. In 
particular, {\sc wyf\_red} performs CCD processing itself if provided with 
bias frames and flat fields. Complete reduction involved several additional 
tasks, the most important of which are detailed below.

\begin{enumerate}
\item Cosmic rays were 
removed automatically from target frames but a manual inspection of
the arc lamp exposures was required to avoid the algorithm mistaking 
arc lines for cosmic rays.

\item Scattered light was subtracted by fitting a smooth function to the 
light found  between the fibres and then interpolating this 
to the regions occupied by the fibre spectra. 

\item The spectra were extracted using an optimal
algorithm which gave more weight to regions of the profile with 
relatively high signal-to-noise ratio (SNR).

\item Each fibre has a different transmission efficiency. 
{\sc wyf\_red} uses the offset blank 
sky spectra to normalise the fibre responses
and also correct for vignetting.

\item Copper-neon arc lamp spectra were used to determine a (fourth order) 
polynomial dispersion function for each fibre. 
The RMS residuals to the arc line fits were typically found to be 
0.005\AA\, and $0.02$\AA\, in the optical and near-infrared regimes 
respectively.

\item All the derived spectra for a target obtained on one night were
co-added.  If an object was observed on both nights then these spectra
were also co-added to find the equivalent widths (EWs) of diagnostic lines, but the
separate spectra from both nights were retained to obtain two
independent radial
velocities (see section~\ref{radvel}).
\end{enumerate}

A sample of the reduced optical and near-infrared spectra are given in
Figs.~\ref{li.montage} and ~\ref{na.montage} respectively. SNR per
pixel was empirically estimated by measuring the residuals to first
order polynomial fits to pseudo-continuum windows. These values are
conservative, as we know that many small molecular lines are probably
present in these regions.  In the $6690-6714$\AA\, region the SNR (per
pixel) ranges from 3 to 20, with a median of 8. Similarly spectra in the
$8175-8210$\AA\, window had a SNR range of 4 to 21 with a median of 9.

One anomaly which requires comment is that during the reduction we found
that two of the fibres allocated to blank sky were seen to 
yield spectra with strong lithium absorption features!
We surmised that there was an error in the translation
table which relates allocated fibre numbers to spectrograph apertures.
This hypothesis was confirmed by examining archival WYFFOS data and shown to 
apply to five pairs of fibres. We have subsequently been able to remove this
source of error from our observations, and mention it here for the benefit
of those who intend to make use of AF2/WYFFOS data (large fibres bundle) 
in the future.

\section{Results}
\subsection{Radial Velocities}
\label{radvel}

\begin{figure}
\includegraphics[width=84mm]{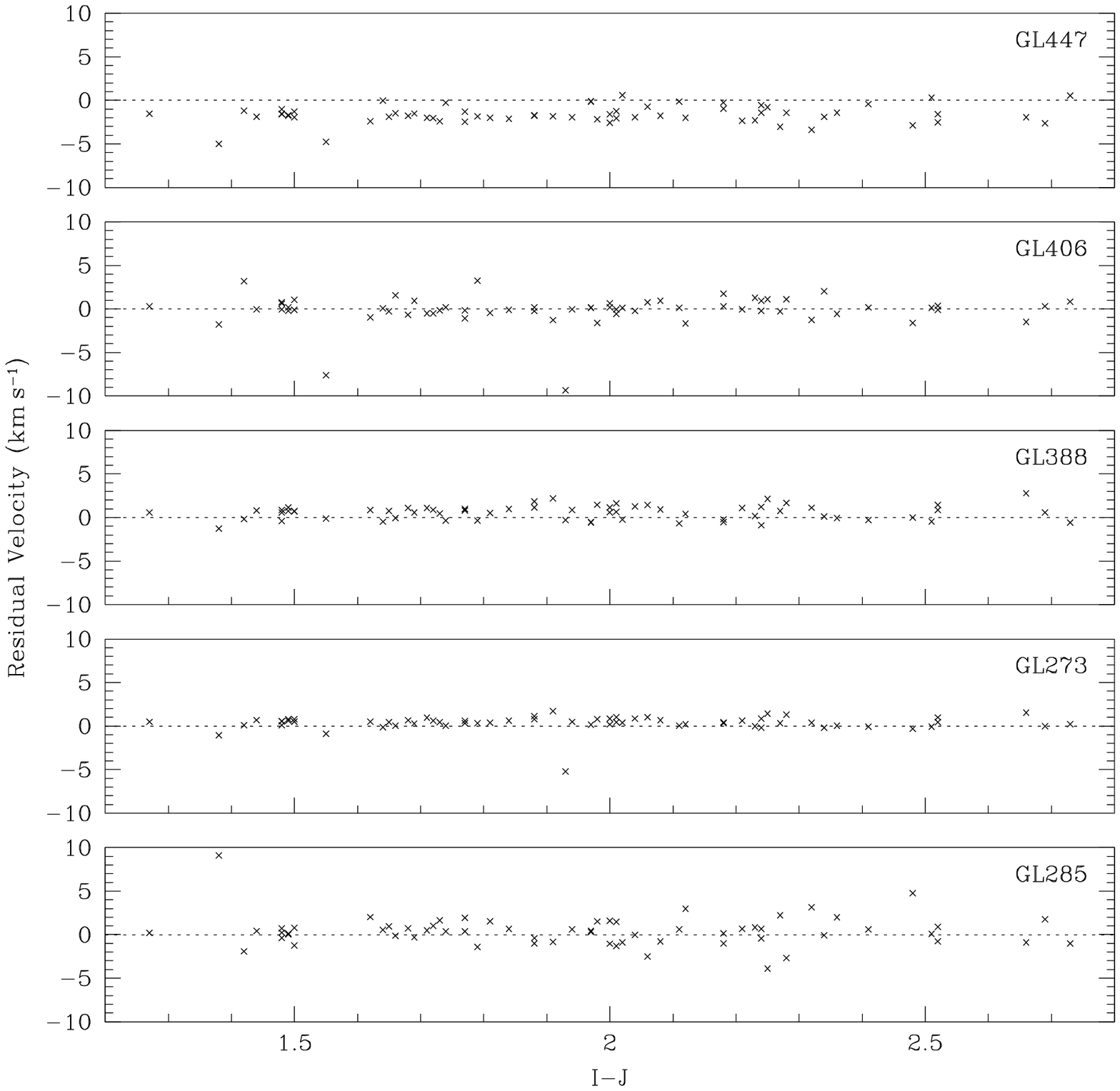}
\caption{Errors introduced into radial velocity measurements by
cross-correlating against a specific template spectrum. Residuals are
the difference between the velocity derived using a specific template
and the mean velocity from all five templates. The dashed line represents
a residual of zero in each panel.}
\label{compare}
\end{figure}

\begin{figure*}
\includegraphics[width=130mm]{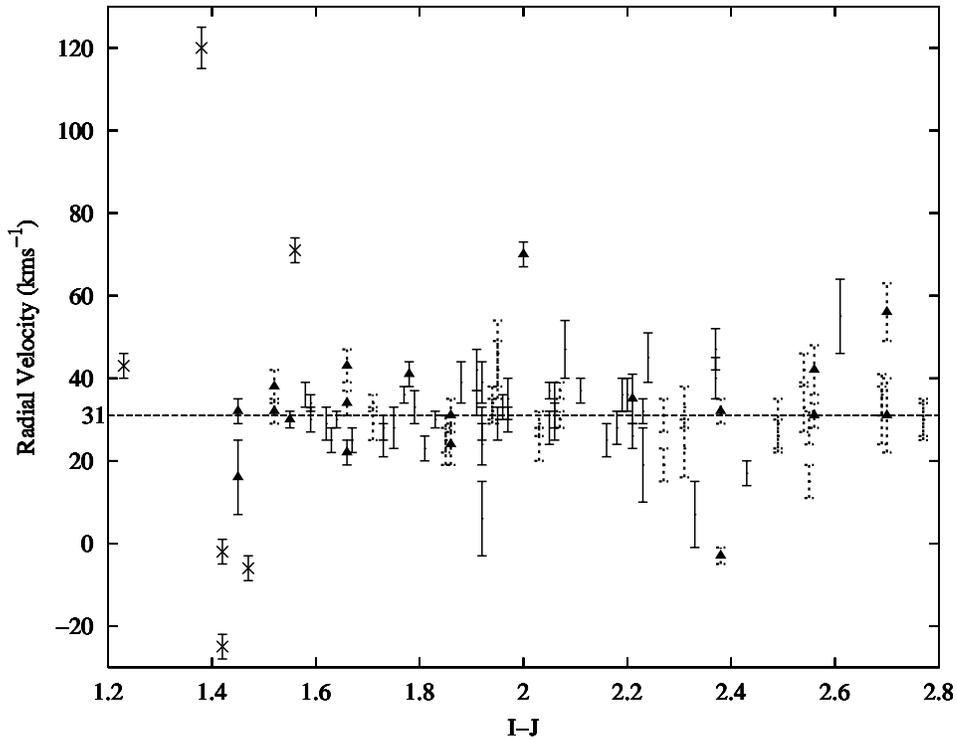}
\caption{Heliocentric radial velocity (RV) as a function of $I-J$
colour. $\sigma$ Orionis
non-members are marked through with crosses, ``grey area'' objects are
triangles (see section 6.1). The 19 objects which were
observed during both runs have the two velocities indicated (shown as dashed
error bars); those for which only one velocity is available are shown as
solid lines. The error bars are $1\sigma$ values, the cluster mean is
indicated by a horizontal line at velocity $31.2$kms$^{-1}$.}
\label{radvelfig}
\end{figure*}

Radial velocities were calculated by cross-correlating the region
8150-8250\AA\,, which included the prominent \sod absorption doublet
(although it is weaker than the standard stars in almost all our target
spectra -- see section~\ref{lithtest}), against five radial velocity
standards (see Table~\ref{rvstands} for details). This spectral window
does contain telluric absorption which could potentially alter derived
radial velocities and \sod EWs.  We removed this contamination to first
order by creating a calibration spectrum from the assumed smooth
continuum of the B3V star we observed. The calibration spectrum was
scaled according to the average airmass of observation (using the {\sc
figaro} program {\sc bsmult}) and used to multiply the target
spectrum. After this correction, very little telluric contamination
could be seen.

\begin{table}
 \caption{M-dwarf spectral type standards. $R-I$ and $I-J$ 
colours are taken from Leggett (1992), heliocentric velocities from Nidever 
et~al. (2002).}
 \label{rvstands}
 \begin{tabular}{@{}lccccc}
 \hline
Object & Sp. &R-I& I-J  & RV$_{H}$ \\
(GL) & Type &    & & (kms$^{-1}$) \\
 \hline
273 &M3.5&1.55  &1.49  & 18.2 $\pm$ 0.1\\
285 &M4.5&1.69  &1.61 & 26.5 $\pm$ 0.3\\
388 &M3.5&1.42  &1.35  & 12.4 $\pm$ 0.1\\
406 &M6&2.18  &2.33  & 19.5 $\pm$ 0.1\\
412b&M6&2.10  & 1.97 & -- \\
447 &M4&1.68  &1.65  & -31.1 $\pm$ 0.1\\
 \hline
 \end{tabular}
\end{table}

The spectra were cross-correlated using the {\sc iraf} package {\sc
fxcor} (Kurtz \& Mink 1998). The spectra were rebinned to a log-linear
dispersion and then continuum subtracted. The ends of each spectrum are
forced to zero with a cosine bell function prior to a fast Fourier
transform being performed. A filter is applied to the Fourier
transforms to limit the frequency window used in the cross-correlation.
The measured cross-correlation lags gave radial velocities which were
then heliocentrically corrected according to the time of observation
and the radial velocity of the standard (see Table~\ref{rvstands}).

To check the success of the cross-correlation process, we investigated the 
effect that using a specific standard star had on:
\begin{enumerate}
\item the cluster mean, 
\item the scatter of objects about this mean.
\end{enumerate}
This was done by cross-correlating all objects against each template spectrum 
after which, a mean velocity per object was calculated. The deviation from 
this mean when each of the templates was used is plotted as a function of 
$I-J$ (a good proxy for spectral type) in Fig.~\ref{compare}.

The top panel suggests that at the time of observation, GL447 had a velocity 
discrepant by about 2\,km\,s$^{-1}$ from that given in Nidever et~al. (2002).
It is also clear that there is more scatter in the RVs determined from
GL 285. In what follows we use the average RV obtained from
cross-correlation versus the spectra of GL273, GL388 and GL406 as there
seems to be little or no dependence of derived radial velocity on
spectral type.

Radial velocity errors were estimated by {\sc fxcor}. We tested their
validity by generating three-hundred template spectra which were
perturbed and degraded to simulate SNRs per pixel of 7, 15 and
25. These simulated spectra were cross-correlated with the original
spectrum from which they were created.  This allowed us to make an
estimate of the error introduced by varying SNRs.  From these
simulations we concluded that the {\sc fxcor} uncertainty estimates
were too small by a factor of $\sim1.75$, which was approximately
independent of SNR.  To remedy this, all the cross-correlation RV
errors were increased accordingly.  An estimate of any remaining error,
as a result of spectral type mis-matches between the object and
template spectra was determined from Fig.~\ref{compare}.  The
mean standard deviation in the RVs derived from each standard of
0.9kms$^{-1}$ was added in quadrature to all RV uncertainties prior to
cluster membership determination. External errors, which must be
considered if comparing our results with other published data, can be
estimated from the deviations of the standard star RVs  from
their published values when their RVs are 
determined by cross-correlation against the other three standards (GL
447 is clearly discrepant and excluded). We estimate that any external
error is less than 1\,km\,s$^{-1}$.

Radial velocities for all objects are listed in Table~\ref{rvmembers}
and displayed in Figure~\ref{radvelfig} as a
function of $I-J$ colour.  Those objects with two independent RV
measurements from different nights are shown with dashed lines.  We
have observed one object (target $60=$ SOri 27) in common with Zapatero Osorio et~al.
(2002), three (targets $39, 55, 75=$ SOri 12, 29, 40) in common with Muzerolle
et~al. (2003), and one (target 44) which appears in Burningham et~al. (2004). 
Table~\ref{compvels} compares the velocities determined
by these authors with the current work.  Our results
concur with the literature in the cases of targets 44, 55 and 60
whereas both targets 39 and 75
display evidence for binarity in the form of a variable radial
velocity. However, for consistency, they are classed as
single from the data in this paper (see section 5.1).

\begin{table}
\caption{Radial velocity comparisons between this work [K04] and values
  in Zapatero Osorio et~al. (2002) [Z02], Muzerolle et~al. (2003) [M03]
and Burningham et~al. (2004) [B04].
Objects are identified as in Table 6.}
\label{compvels}
\begin{tabular}{@{}lcccc}

\hline
\multicolumn{1}{l}{}&
\multicolumn{3}{c}{Radial Velocity (kms$^{-1}$)}\\

\multicolumn{1}{l}{ID}&
\multicolumn{1}{c}{Z02}&
\multicolumn{1}{c}{M03}&
\multicolumn{1}{c}{B04}&
\multicolumn{1}{c}{K04}\\
\hline

\multicolumn{1}{c}{39}&
\multicolumn{1}{l}{}&
\multicolumn{1}{l}{29.8 $\pm$ 0.7}&
\multicolumn{1}{l}{}&
\multicolumn{1}{l}{37 $\pm$ 2}\\

\multicolumn{1}{c}{44}&
\multicolumn{1}{l}{}&
\multicolumn{1}{l}{}&
\multicolumn{1}{l}{31.1 $\pm$ 4.1}&
\multicolumn{1}{l}{34 $\pm$ 3}\\

\multicolumn{1}{c}{55}&
\multicolumn{1}{l}{}&
\multicolumn{1}{l}{27.1 $\pm$ 1.6}&
\multicolumn{1}{l}{}&
\multicolumn{1}{l}{28 $\pm$ 4}\\

\multicolumn{1}{c}{60}&
\multicolumn{1}{l}{35.5 $\pm$ 10}&
\multicolumn{1}{l}{}&
\multicolumn{1}{l}{}&
\multicolumn{1}{l}{47 $\pm$ 5}\\

\multicolumn{1}{c}{75}&
\multicolumn{1}{l}{}&
\multicolumn{1}{l}{32.5 $\pm$ 3.3}&
\multicolumn{1}{l}{}&
\multicolumn{1}{l}{55 $\pm$ 9}\\

\hline

\end{tabular}
\end{table}

\subsection{Lithium \& Sodium}\label{lithtest}

\begin{figure}
\includegraphics[width=84mm]{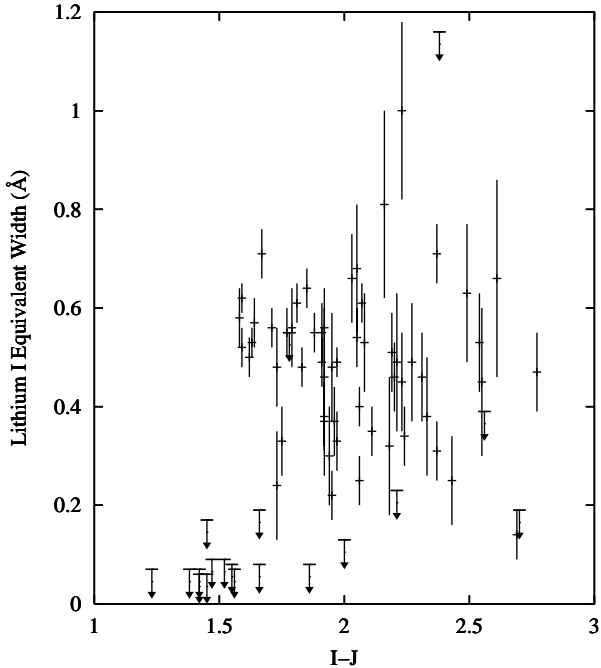}
\caption{Lithium I (6708\AA) equivalent width versus $I-J$ colour for our 
objects. Where no lithium line was observed, 2$\sigma$ upper limits are 
shown.}
\label{li.ij}
\end{figure}

\begin{figure}
\includegraphics[width=84mm]{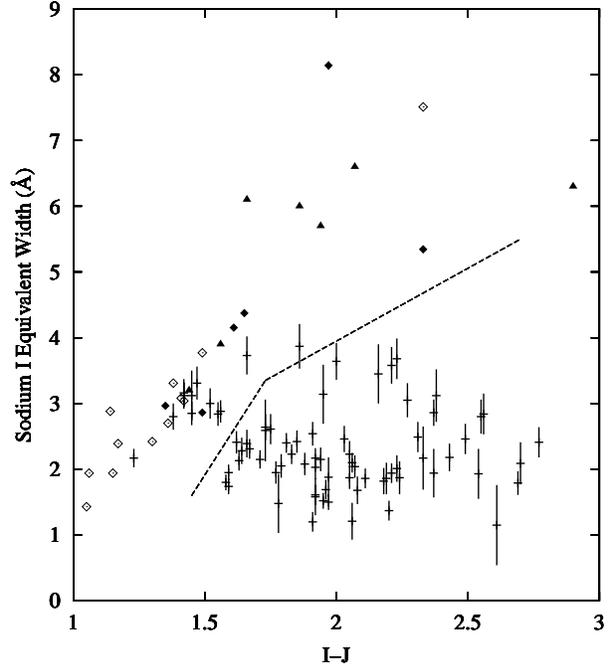}
\caption{Sodium I (8183,8195\AA) equivalent widths versus $I-J$
colour. Filled diamonds indicate
our standard star measurements, open diamonds represent M-dwarfs from Xu
(1991) and triangles are M-dwarf measurements from Mart{\' \i}n et
al. (2004). Objects lying below and to the right of the dashed line are
considered to have gravities lower than field M-dwarfs.
}
\label{na.ij}
\end{figure}

The EWs of both the \lith 6708\AA\ line and \sod near IR
doublet at 8183, 8195\AA\ can be used as indicators of age and hence
membership in samples of VLMS and BDs (see section~\ref{membership}).
A low-order polynomial function was used to normalise the spectra using
pseudo-continuum regions around both of these features (and estimate
the SNR). The absorption lines were modelled as 
simple Gaussian functions (two Gaussians in the case of the
\sod doublet) and the results integrated to estimate the EWs.  
1-sigma uncertainties in the EW measurements were estimated as
\begin{equation}
\label{snrequation}
\delta EW = \frac{\sqrt{{\rm FWHM}\times p}}{SNR}\, ,
\end{equation}
where FWHM and $p$ are the full width half maxima and pixel size
respectively, in units of \AA. The pseudo-continuum
regions either side of the lines are large enough that the {\it
  statistical} errors due to the uncertainty in the continuum level are
negligible. The \sod EWs for the field M-dwarf (RV)
standards were also measured in the same way. 

Figures~\ref{li.ij} and \ref{na.ij} show the lithium and sodium EWs
plotted as a function of $I-J$ colour.  Where no
significant \lith feature was seen, 2-sigma upper limits to the
equivalent width have been calculated based on a the median line FWHM
and the SNR of the spectrum in question.  This allows the distinction
to be made between poor SNR spectra which may be masking the lithium
line and objects which certainly do not possess the line.
It is worth emphasizing that although bright forbidden lines of S\,{\sc ii}
produce a very noisy sky subtraction at 6716/6731\AA, the sky background at 
$\pm 5$\AA\ from the \lith 6708\AA\ line is low and well behaved (see Fig.~3).

Our survey includes a handful of objects which have had their \lith and
\sod features measured by other authors (B\'ejar et al. 1999;
Zapatero-Osorio et al. 2002; Muzerolle et al. 2003). A comparison of
the EWs is presented in Table~\ref{bejartable}. There is reasonable
agreement between our \lith EWs and those found in the high resolution
spectra of Muzerolle et al. (2003). However, there are some clear
discrepancies between our EW values and those found in lower
resolution spectra by B\'ejar et al. (1999) and Zapatero-Osorio et al. (2002).

\begin{table}
\caption{Sodium I and Lithium I equivalent widths for objects common to
B\'ejar et~al. (1999) [B99], Zapatero Osorio et~al. (2002) [Z02],
Muzerolle et~al. (2003) [M03], Burningham et~al. (2004) [B04] 
and the current work [K04]. Objects are identified as in Table 6.
Uncertainties for the equivalent widths are $\pm1$\AA\, for the [B99] sample, 
$\pm0.09$\AA\, for the [Z02] point and $\pm0.39$\AA\, for the [B04] object.}

 \label{bejartable}
 \begin{tabular}{@{}lcccccc}
\hline
 \multicolumn{1}{l}{}&
 \multicolumn{3}{c}{EW(\sod)} &
 \multicolumn{3}{c}{EW(\lith)} \\

 \multicolumn{1}{l}{ID}&
 \multicolumn{1}{c}{B99}&
 \multicolumn{1}{c}{B04}&
 \multicolumn{1}{c}{K04}&
 \multicolumn{1}{c}{Z02}&
 \multicolumn{1}{c}{M03}&
 \multicolumn{1}{c}{K04}\\

\hline

 \multicolumn{1}{c}{39}&
 \multicolumn{1}{c}{2.5}&
 \multicolumn{1}{c}{}&
 \multicolumn{1}{c}{2.23$\pm$0.20}&
 \multicolumn{1}{c}{}&
 \multicolumn{1}{c}{0.6}&
 \multicolumn{1}{c}{0.54$\pm$0.06}\\

 \multicolumn{1}{c}{44}&
 \multicolumn{1}{c}{}&
 \multicolumn{1}{c}{2.65}&
 \multicolumn{1}{c}{2.54$\pm$0.18}&
 \multicolumn{1}{c}{}&
 \multicolumn{1}{c}{}&
 \multicolumn{1}{c}{}\\

 \multicolumn{1}{c}{55}&
 \multicolumn{1}{c}{2.2}&
 \multicolumn{1}{c}{}&
 \multicolumn{1}{c}{1.82$\pm$0.20}&
 \multicolumn{1}{c}{}&
 \multicolumn{1}{c}{0.6}&
 \multicolumn{1}{c}{0.32$\pm$0.14}\\

 \multicolumn{1}{c}{60}&
 \multicolumn{1}{c}{$\leq1.0$}&
 \multicolumn{1}{c}{}&
 \multicolumn{1}{c}{2.86$\pm$0.20}&
 \multicolumn{1}{c}{0.74}&
 \multicolumn{1}{c}{}&
 \multicolumn{1}{c}{0.31$\pm$0.06}\\

 \multicolumn{1}{c}{68}&
 \multicolumn{1}{c}{$\leq1.0$}&
 \multicolumn{1}{c}{}&
 \multicolumn{1}{c}{3.68$\pm$0.31}&
 \multicolumn{1}{c}{}&
 \multicolumn{1}{c}{}&
 \multicolumn{1}{c}{}\\

 \multicolumn{1}{c}{75}&
 \multicolumn{1}{c}{$\leq1.0$}&
 \multicolumn{1}{c}{}&
 \multicolumn{1}{c}{1.15$\pm$0.60}&
 \multicolumn{1}{c}{}&
 \multicolumn{1}{c}{0.5}&
 \multicolumn{1}{c}{0.66$\pm$0.20}\\

\hline

\end{tabular}
\end{table}

\subsection{H$\alpha$}

\begin{figure*}
\includegraphics[width=110mm]{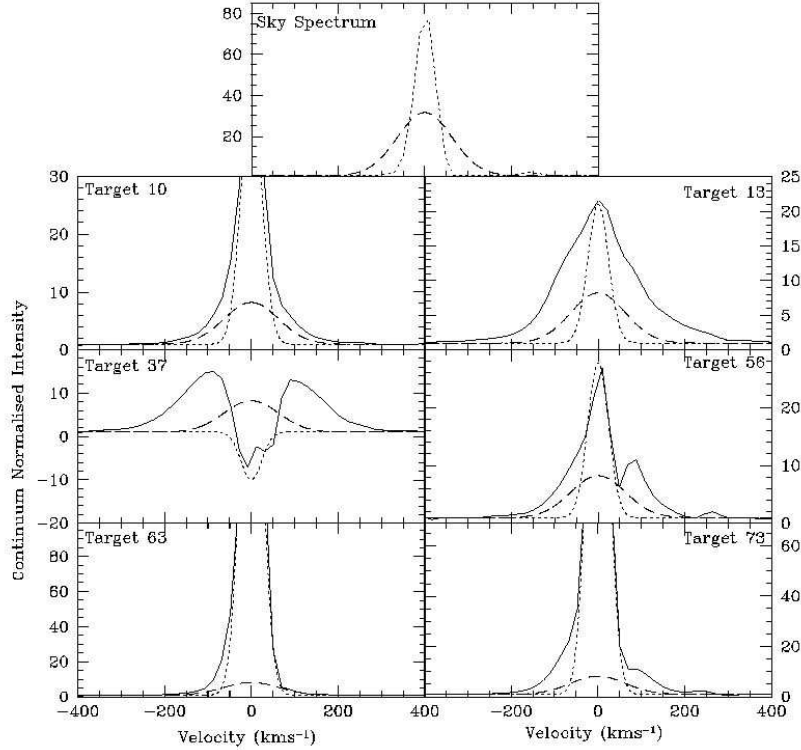}
\caption{H$\alpha$ velocity profiles of candidate accreting
objects. The top panel shows a typical sky profile (dotted line), plus
a Gaussian profile (dashed line) with equivalent width $25$\AA\ and a
full width at 10 per cent of peak of $\pm 135$\,km\,s$^{-1}$. Each
subsequent panel also shows the observed H$\alpha$ line (with continuum
normalised to unity), with an
overplotted sky spectrum (scaled to the peak H$\alpha$ flux)
and a minimal accretion profile (the dashed Gaussian) to illustrate
that the profile is (a) broader than any possible contribution from the
sky and (b) broader or stronger than the minimum expected profile from
any accreting object.}
\label{wide}
\end{figure*}

Strong H$\alpha$ emission is a characteristic of young stars, both as a
result of magnetic chromospheric activity and from continuing
accretion.  Our attempts to study H$\alpha$ in the
$\sigma$ Ori candidates are hampered by the presence of strong, spatially varying
H$\alpha$ emission that can be clearly seen in our $R$-band images and
is presumably associated with hot gas in the line of sight towards $\sigma$~Ori. 
The signal from the background H$\alpha$ is very strong and, in
our 2.7 arcsec diameter fibres, swamps any H$\alpha$ signal from the
targets, making a measurement of the H$\alpha$ EW impossible. However,
the background emission is unresolved in our sky spectra, with a FWHM
of about 40-50\,km\,s$^{-1}$. White and Basri (2003) have demonstrated
that H$\alpha$ emission lines formed by accretion processes should be
much broader than this. Classical T-Tauri stars typically exhibit
H$\alpha$ lines with a full-width at 10\% of the peak emission of
$>270$kms$^{-1}$. So, whilst we cannot measure the strength of the
H$\alpha$ line, we can certainly search for emission wings
extending to more than $\pm 135$\,km\,s$^{-1}$ from the line centre,
which are uncontaminated by the background H$\alpha$ emission.

The H$\alpha$ EW expected from accreting objects which demonstrate
extended H$\alpha$ emission with spectral types M3-M7 (the approximate
range of our sample) is greater than 25\AA\ and more usually several
times stronger (e.g. see White \& Basri 2003; Muzerolle et
al. 2003). Given the dispersion of our spectra and assuming the
H$\alpha$ profile is pseudo-Gaussian it is easy to show that such a
line would produce a signal of strength $>0.7$ times the continuum
level at $\pm 135$\,km\,s$^{-1}$ from the line centre. Such a signal
should be easily detectable if present, even in our poorest spectra.

Analysis of our spectra yields five objects which clearly exhibit
extended H$\alpha$ emission on this basis (see Fig.~8) and a further
one (target 63) which has a marginally broadened line which may simply be
a case of poor sky subtraction or the result of small changes in the
instrumental dispersion profile over the detector.  These objects are
marked with an ``a'' in the last column of Table~\ref{rvmembers}.

\section{Membership}

\label{membership}

Determining the membership status of candidates can be a subjective
process largely dependent on the specific criteria set. This section
details the three diagnostic tools we used to assign membership of the
$\sigma$ Orionis cluster.

\subsection{Radial Velocities}
Our initial indicator of cluster membership was based on comparing
individual RVs to the weighted mean cluster velocity.
Targets lying more than $3\sigma$ away from this mean were flagged as
likely non-members (or binaries) 
and excluded from subsequent recalculations.  In the
$19$ cases where two RVs were available for a given
object, the average of the two was used in computing the ensemble
velocity.  The result of this iterative process was that $66$ objects
were designated as $\sigma$ Orionis RV cluster members,
lying within $3\sigma$ of a cluster mean velocity of $31.2\pm
0.2$\,km\,s$^{-1}$ (which has an additional external error of up to
1\,km\,s$^{-1}$).

Our mean RV for the cluster agrees reasonably well with the systemic
velocity found for $\sigma$~Ori by Morrell \& Levato (1991) of
27\,km\,s$^{-1}$; the value of $29.1$\,km\,s$^{-1}$ listed for
$\sigma$~Ori by Evans (1967); the peak in the RV distribution of
low-mass PMS candidates around $\sigma$~Ori at 25-30\,km\,s$^{-1}$
found by Walter et al. (1998) and at 29.5\,km\,s$^{-1}$ by
Burningham et al. (2004). It is somewhat lower than the mean of
37.3\,km\,s$^{-1}$ reported for a group of candidate low-mass members
by Zapatero-Osorio et al. (2002).

The iterative clipping process has the potential to exclude candidate
members which are cluster binary systems with a small enough
separation to introduce RV variations that move them away from the
cluster mean. If other information suggests strongly enough that
candidates are indeed $\sigma$ Ori association members, then a single
discrepant RV measurement can be sufficient to identify a short period,
single-lined spectroscopic binary.  So, all objects failing the RV test
must be examined closely for any convincing signs of membership from
other indicators.

From these considerations we have 10 RV
non-members in our sample,
with heliocentric RVs ranging from -25 to 120\,km\,s$^{-1}$. If we
approximate our RV selection to a band of width $\sim25$\,km\,s$^{-1}$
centred on the cluster mean and assume that contaminating RV
non-members are spread evenly in RV space then we would only expect of order
1-2 contaminating non-members to be left among our RV-selected members.

\subsection{Sodium}
Each of our $\sigma$ Ori candidates exhibits the gravity sensitive
$8183,8195$\AA\, sodium doublet in their near-infrared spectra. The
strength of this doublet is gravity sensitive (see Schiavon et
al. 1997). It is 2-3 times stronger in cool M dwarfs (M4 and later)
than in PMS stars with age $<10$\,Myr and similar effective temperature
(e.g. Kirkpatrick et al. 1991; Mart{\' \i}n, Rebolo \& Zapatero-Osorio
1996; Mart{\' \i}n, Delfosse \& Guieu 2004).  As demonstrated in Figure
\ref{na.ij}, this test may only be conclusive for objects with
$I-J>1.6$. Warmer objects in the $\sigma$ Ori cluster will have
gravities only slightly lower than field objects with the same colour.
We define an arbitrary empirical boundary, plotted on Figure
\ref{na.ij} with a dashed line, below which we are confident that a low
\sod EW indicates a gravity consistent with PMS status.  64 candidates
lie below this line and $12$ above.

The \sod EW is also very weak in giant stars ($\leq 1$\AA\ according
to Schiavon et al. 1997), but contamination of our
sample by M-type giants is highly improbable. An M-giant at the
appropriate location on the CMD would have an absolute
$I$ magnitude $<-2$ and hence would have to be at distances
$>25$\,kpc. As the galactic latitude and longitude of $\sigma$ Ori are
$l=206.8^{\circ}$, $b=-17.3$, such objects would be more than
7.4\,kpc below the galactic disc, where their spatial density is known
to be essentially zero (Branham 2003).

\subsection{Lithium}
The presence of undepleted lithium should be an excellent indicator of stellar
youth, because it is burned in the fully convective interiors of
low-mass PMS stars once their cores reach a threshold temperature.  We
would expect Li to start being depleted by more than a factor of 100
among contaminating field VLMS at ages between 20\,Myr (at 0.35\,$M_{\odot}$)
and 120\,Myr. Field
BDs with masses below about 0.06\,$M_{\odot}$ never deplete Li, but
these would have colours redder than $I-J>2.8$ (the reddest in our
sample) for ages $>140$\,Myr (Baraffe et al. 1998). According to
tables of EW versus Li abundance
presented by Zapatero-Osorio et al. (2002), which have
appropriate temperatures for the sample considered here (at 5\,Myr
$1.4<I-J<2.8$ corresponds to $3500>T_{\rm eff}>2650$\,K -- Baraffe et
al. 1998), undepleted Li should result in a \lith 6708\AA\ line EW of
about 0.5-0.6\AA\, falling to 0.2-0.3\AA\ after a factor of 100
depletion and becoming much weaker shortly after that. Thus the
detection of \lith 6708\AA\ consistent with an EW of 0.2\AA\ or more should
be a very strong indicator of youth in the hottest stars of our sample
(implying ages $<20$\,Myr), but is less constraining (ages $<140$\,Myr)
for the coolest stars in our sample. Only a small fraction ($\simeq
0.1-1$ per cent) of any contaminating foreground field stars might be
expected to show evidence for Li at this level.

A caveat to this picture might be if there are remnants from an older
generation of star formation in the Orion region which might show
Li. However, these older stars would have to lie in the foreground (to
appear in the CMDs in the correct place) and would also have to satisfy
the RV tests to be included as candidate members.  We have to bear in
mind that potential binary members exhibiting Li, and a weak \sod
feature but with a discrepant RV could also be members of a young
(10-100\,Myr) foreground population.

Figure \ref{li.ij} shows the $57$ objects with available
equivalent widths (and photometry) along with $19$ calculated upper limits. 
We decide to rule out membership for any star which has
EW[Li]$<0.2$\AA\, at the 2-sigma level. Above this level the presence
of Li is taken to be very strong evidence of cluster membership and if
the Li measurement upper limit is inconclusive then the evidence from
the other two membership indicators is used.  However, we find a number
of objects that display weak or absent Li but which appear to have
low-gravity and are RV members and we will return to these in Section 6.

\section{Discussion}
\subsection{Membership Indicators\label{membindic}}
Using three membership tests means that candidates can be placed into
a number of categories.

\begin{enumerate}

\item Most candidates ($55/76$) are 
RV members with weak gravity and display evidence 
of significant lithium in their atmospheres. These are clearly the strongest cases for 
being true $\sigma$ Ori members. All but one of the strong H$\alpha$ emitters/accretors
belong in this category.


\item There are 3 candidates (targets 27, 40 and 73 -- an H$\alpha$ emitter/accretor) which have an inconclusive upper limit to their Li EW, but which are
RV members and display low gravity.

\item There are 3 candidates that display a low gravity, are RV
  non-members and either show Li (targets 46 and 74) or have an upper
  limit to their Li consistent with the presence of significant Li
  (target 72). These are possible binary members of the
  cluster. In the case of target 72, the RV is
  demonstrably variable.

\item There are 8 candidates which have no Li, are RV members and have a
  Na EW that is either inconclusive (targets 29, 33, 34, 45, 53 and 59) or
  indicates a low gravity (targets 25 and 76).

\item There is 1 candidate (target 52) which shows no evidence for Li, has a
  discrepant RV yet a weak Na EW, indicative of low gravity.

\item There are 6 candidates (targets 7, 12, 26, 28, 30 and 38) that have very
  low upper limits to their Li EW, inconclusive Na EW and highly
  discrepant RVs. These are all considered definite non-members
\end{enumerate}

In summary we have found what we believe are 57 firm members of the
$\sigma$~Ori cluster (those displaying Li), of which 2 are probable short
period binary systems. We have also identified 6 objects as clear non-members
of the cluster. 
Inevitably some of our
classifications are arguable, but there is sufficient information in
Table~\ref{rvmembers} for the reader to
generate their own classifications.

Thirteen of our targets (an uncomfortably large number) lie in a grey
area where we find it difficult to assign membership, either because
they show low-gravity but discrepant RVs, low-gravity but no Li, or the
right RV but no Li.  Only 2 out of these 13 show discrepant RVs, one of
which is definitely an RV variable. Given the relatively narrow range
of RVs occupied by members and the spread of velocities occupied by the
10 (out of the total 76) targets with discrepant RVs, we would only
have expected 1-2 non-members to have the ``correct'' RV. This presents
us with a puzzle, because it suggests that the majority of the 11
``grey area'' objects with the ``correct'' RV should be cluster
members.  Three of these objects have an upper limit to the Li EW which
could be consistent with negligibly depleted Li, but 8 do not, and of
these, 2 (targets 25 and 76 -- see Fig.~\ref{li.montage}) have a small Na EW
that would seem to unambiguously classify them as low gravity PMS
objects. This in turn suggests there may be a problem in using a simple
EW[Li] threshold to discriminate between cluster objects and
non-members.  On the other hand, that 2 out of 13 grey area objects are
RV non-members contrasts with only 2 out of 57 Li-rich RV
non-members. So unless short period binarity somehow has a tendency to
reduce EW[Li], it is probable that one or more of these 2 are not
cluster members, despite appearing to be low-gravity objects (see also
section~6.4); in which case we have the problem that EW[Na] may not be
a foolproof membership indicator.

\subsection{A spread in Li equivalent widths?}

\begin{figure}
\includegraphics[width=84mm]{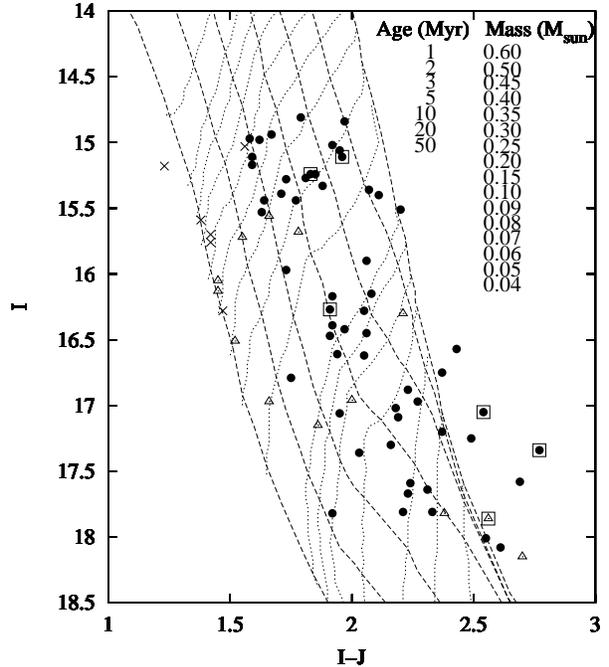}
\caption{$I$,$I-J$ colour-magnitude diagram. Symbols have the same
  meaning as in Fig.~1 with open squares around those objects found to
  have accretion. Evolutionary 
tracks for masses $0.04-0.6$M$_{\sun}$ (dotted lines) and isochrones for ages
$1-50$Myr (dashed lines) are calculated using 
the models of Baraffe et al. (1998). The isochrones range from 1\,Myr
(far right) to 50\,Myr and the evolutionary tracks are shown for masses
from 0.6\,$M_{\odot}$ (at the top) to $0.04\,M_{\odot}$.}
\label{cmd}
\end{figure}

We have observed a large spread in the measured values of
EW[Li]. The spread appears to be confined to those cluster candidates
with $I-J>1.7$ (see Fig.~\ref{li.ij}) and cannot be explained with the
quoted experimental errors -- many of our spectra are of reasonably
high quality and we believe the {\em statistical} errors in the EW
values are quite robust, if not pessimistic. In particular there appear
to be a substantial fraction of the cooler stars which have EW[Li]$\sim
0.3$\AA.  In addition there are also the ``grey area'' objects, the
majority of which are likely to be members but most of which appear to
exhibit no Li (including three with a definite PMS gravity). 

It also seems that the average EW[Li]
declines with increasing $I-J$ from about 0.6\AA\ to 0.5\AA\ across our
full $I-J$ range. We mention this because the tables of EW versus Li abundance
presented by Zapatero-Osorio et al. (2002) show instead a
slight increase in the expected EW[Li] for an undepleted Li abundance, 
of about the same magnitude.  A
plausible explanation for the decreasing average EW[Li] with $I-J$ in
our sample arises from the definition of the pseudo-continuum used to
estimate EW[Li]. The continuum regions we have used cover a
rather wide range around the Li feature, so they must surely contain
molecular absorption features which become stronger at cooler
temperatures. This would depress the pseudo-continuum leading to lower
estimates of EW[Li].

A real Li abundance spread could be caused by depletion as Li is burned
in the cores of contracting young low mass objects. As explained in
section 5.3, factors of 100 depletion in Li are probably needed to
reduce EW[Li] to $\leq0.3$\AA, which does not occur (according to the
models of Baraffe et al. 1998) until ages of 20-120\,Myr.  Other
models, such as those of D'Antona \& Mazzitelli (1997) or Siess, Dufour
\& Forestini (2000), predict similar if not slightly larger timescales
for Li depletion. Figure~\ref{cmd} shows the $I$ versus $I-J$ colour
magnitude diagram for our targets together with isochrones from the
Baraffe et al. (1998) models. The models are adjusted for the Hipparcos
distance to $\sigma$~Ori of 352\,pc and suggest a spread of ages
between $<1$ and 20\,Myr for our Li-confirmed cluster candidates.  Only
some of the oldest, most massive stars in our sample could have
suffered significant Li depletion if these ages are taken
literally. However, the ``oldest'' Li-poor cluster candidates among the
bright stars in our sample also have discrepant RVs and so are almost
certainly cluster non-members. In any case the apparent age spread present in
the diagram is likely to be a vast overestimate -- photometric
variability, binarity and for some stars, the presence of discs, can
perhaps entirely explain such spreads (e.g. Hartmann 2001). We conclude
that Li-depletion among members of the $\sigma$ Ori cluster is unlikely
to be the cause of spreads in EW[Li] unless current PMS evolutionary
models have very serious flaws.

An alternate possibility is that some of our cluster candidates are
genuinely depleted of Li because they {\em are} older and are not
cluster members.  They could belong to a foreground population of older
stars that are at an appropriate distance to place them close to the
$\sigma$~Ori cluster locus in CMDs (see Alcal\'{a} et al. 2000; Walter
et al. 2000). Such
objects would need to have a similar RV to the cluster (which is quite
different to the general field population), be old enough to have
depleted a factor of 100 or more of their initial Li and be yet be
young enough to exhibit weak Na lines. There may be a vanishingly small
range of ages that can satisfy these requirements and whilst a slightly
older population (the Orion OB1a subgroup; age$~11$\,Myr) does exist in
some parts of the Orion star forming region, it is at a similar
distance and there is no evidence of this older component in the
direction of $\sigma$~Ori (Sherry 2003). To conclude, we find this
possibility unlikely.

A second class of explanations are those in which the spread is
apparent in the EWs because of deficiencies in the atmospheric analysis
or the data reduction. The EW-abundance relationships we are using to interpret
our results show that the EW[Li] versus $T_{\rm eff}$ relationship is
quite flat over the $T_{\rm eff}$ range considered here. It is
therefore unlikely that uncertainties in the colours play any
significant role in the EW[Li] scatter. Likewise, temperature
inhomogeneities on the stellar surfaces caused by magnetic fields are
are also unlikely to have much effect. Some photospheric lines can be
veiled in young stars that are undergoing accretion, but the majority
of stars in our sample are accreting at a rate
$M<10^{-11}M_{\odot}$/year (see section 6.5) which is insufficient to cause such an
effect (Muzerolle et al. 2003). Also, the 5 or 6 objects which do show
evidence for ongoing accretion have EW[Li] values of 0.32\AA\ to 1.19\AA,
which are not systematically low compared with the rest of the sample.

We have also considered
possible data reduction problems such as an uncertain and perhaps systematically
underestimated scattered light or sky contributions which would
preferentially affect the fainter targets. The level of scattered light
in the spectrograph is simply too small to be of concern here. The sky
contribution is more significant, but we estimate that it is subtracted
with a precision of better then 5 per cent based on the scatter in the
fibre-to-fibre relative efficiencies determined from three successive
offset sky exposures. We have confirmed that this would not
significantly increase the quoted EW[Li] uncertainties 
for any of our targets.

Finally we note that a spread in EW[Li] among VLMS and BDs has been
seen before, although not extensively discussed. Both J\"orgens \&
Guenther (2001) and Natta et al. (2004) have reported significant
EW[Li] spreads of $<0.09$ to 0.63\AA\ and 0.2 to 0.5\AA\ respectively,
among a small group of very low mass objects in the young Chamaeleon~I
cluster. It is also notable that a subset of these objects were
observed earlier by Neuh\"auser \& Comer\'on (1999), albeit at lower
resolution, who found quite different EW[Li] for some objects.

We can reach no firm conclusion here. The EW[Li] spread demands further
attention with better spectra, if only to rule out the possibility of
major problems with the current evolutionary models. Our results cast
doubt on any selection technique which rules out membership of young
clusters solely on the basis of a small EW[Li].

\subsection{Contamination and the mass function}

Previous work on the $\sigma$ Ori cluster has attempted to estimate the
mass function of VLMS and BDs using photometrically
selected samples, finding $dn/d\log m \propto m^{-\alpha}$, with $\alpha =
-0.2\pm 0.4$ for $0.01<m<0.2\,M_{\odot}$ (B\'{e}jar et al. 1999, 2001).
Only a minority of this sample have been followed up spectroscopically
and in most cases, this spectroscopy is limited to obtaining a spectral
type and noting the presence of moderate to strong H$\alpha$ emission
in low resolution (20\AA\ FWHM) spectra. We contend that such an
approach may not yield reliable results. As we discussed earlier, the main source of any
contamination in the sample will be late-type foreground M dwarfs that
have spectral types, colours, and in many cases H$\alpha$ emission,
that are similar to members of $\sigma$ Ori. A very large H$\alpha$ EW
of perhaps 20-40\AA\ or more {\em is} associated with accretion and hence
youth, but is present in only a small fraction of $\sigma$ Ori members
(see section~\ref{discs}).
Measurement of radial velocities, Li and weak gravity using Na and K
line strengths will be much more reliable, but is only possible with
higher resolution spectra.

Arguments based upon the space density of field M dwarfs, have
suggested that any contamination in a photometrically selected sample
may be quite low (see B\'{e}jar et al. 1999 and below). Our spectroscopy and membership
classification partially reinforce this view. We have found at least 57
(and probably closer to 64) likely members
from a sample of 76 selected {\em only} on the basis of their $I$
magnitudes and $R-I$ colours. Cluster members are found close to the
edge of the sample distribution in both the $I$ vs $R-I$ and $I$ vs
$I-J$ CMDs, suggesting that the photometric selection criteria may have
been too restrictive to include {\em all} the possible members of
$\sigma$ Ori. RV measurements of another faint sample of
$\sigma$~Ori candidates spread over a wider region of the $I$ vs $R-I$
CMD by Burningham et al. (2004) suggest that few members lie outside
the region we have considered. However, because of this uncertainty and 
because we have not observed a complete sample of photometric
candidates over our restricted field of view, we limit ourselves
in this paper to asking whether a mass function derived solely by photometric selection
is likely to be significantly affected by non-member contamination?

The definite non-members we have found are at the brighter end of
the sample ($14.8<I<16.3$) -- where the contamination level is 15-25 per
cent (depending on the status of the ``grey area objects''). 
Based solely on their low EW[Na], it is possible that 
{\em all} of the fainter objects are members, but the contamination
level could be as high as 20 per cent if all the ``grey area'' objects
turn out to be non-members. 
Contamination at this level would have minimal impact on the value
of $\alpha$, especially when compared to the effect of statistical, distance and
model-dependent uncertainties.

To check our estimates of contamination we have simulated an $I$ versus
$R-I$ diagram populated by field dwarfs. The details of this
simulation, which uses the local field dwarf luminosity function
and space density, are presented in Jeffries et al. (2004). We
simulated a 0.5 square degree field towards $\sigma$~Ori, similar to
that area from which our spectroscopy targets were chosen. An $I$
versus $R-I$ diagram was generated for objects up to 500\,pc
distant. We then selected objects with $14.8<I<18.2$ and in a strip 0.6
magnitudes wide in $R-I$ and centred on the 5\,Myr isochrone shown in
Fig~1.  We found that 41 simulated M-dwarfs lay within this strip.  In
the real photometry dataset there are 115 objects in the same strip,
of which 74 were observed spectroscopically and $57-70$ are found to be
members depending on the status of ``grey area'' objects. Hence we
would expect to have observed about 25 non-members in our spectroscopic
sample, compared to the 4-17 we have found. We consider this
reasonable agreement, although any discrepancy might be explained by a
tendency for us to prefer to allocate fibres to spectroscopic targets
which lay close to the 5\,Myr isochrone. Most of the contaminants will
lie on the blueward side of the selected strip. Indeed only 2 simulated
contaminants lay on the red side of the 5\,Myr isochrone and Fig.~1
shows that all the confirmed non-members and all but one of the ``grey
area'' objects lie blueward of the 5\,Myr isochrone.

It is interesting to note that had we applied a strict age$\leq 20$\,Myr
criterion using the $I$ vs $I-J$ diagram shown in Fig.~\ref{cmd}, we
would have excluded all of the definite non-members from our
sample. The 4 ``grey
area'' objects that would be excluded on this basis have no Li, have an ambiguous
EW[Na] and in two cases an RV only marginally consistent with cluster
membership. Selection on the $I$ versus $I-J$ diagram seems therefore to have the
potential for being more discriminating in choosing a sample of genuine
cluster candidates.

%
%

\subsection{Binarity}

\begin{figure}
\includegraphics[width=84mm]{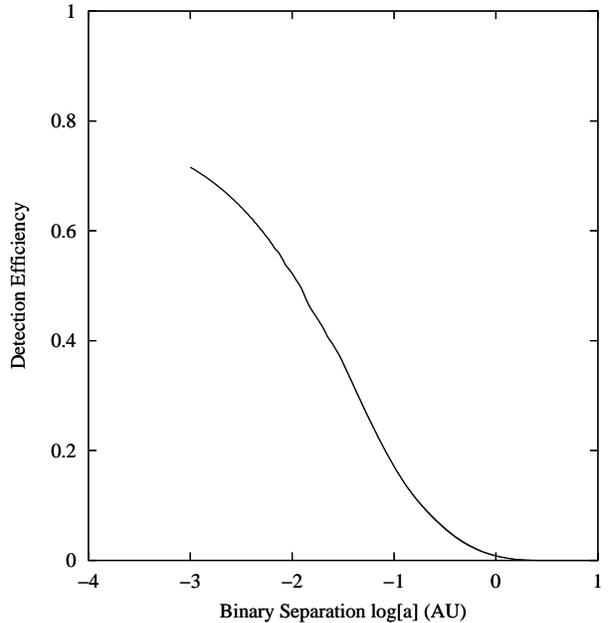}
\caption{The efficiency with which we are able to detect binaries as a
  function of separation. The solid curve is averaged over the 57 objects in
  sample~A (see text). A similar curve for sample~B is almost indistinguishable.
  We have assumed circular orbits, random inclination
  angles, a flat mass ratio distribution for $0.2<q<1$ and primary
  masses taken from Fig.~\ref{cmd}.
}
\label{subfig1}
\end{figure}

\begin{figure*}
\includegraphics[width=120mm]{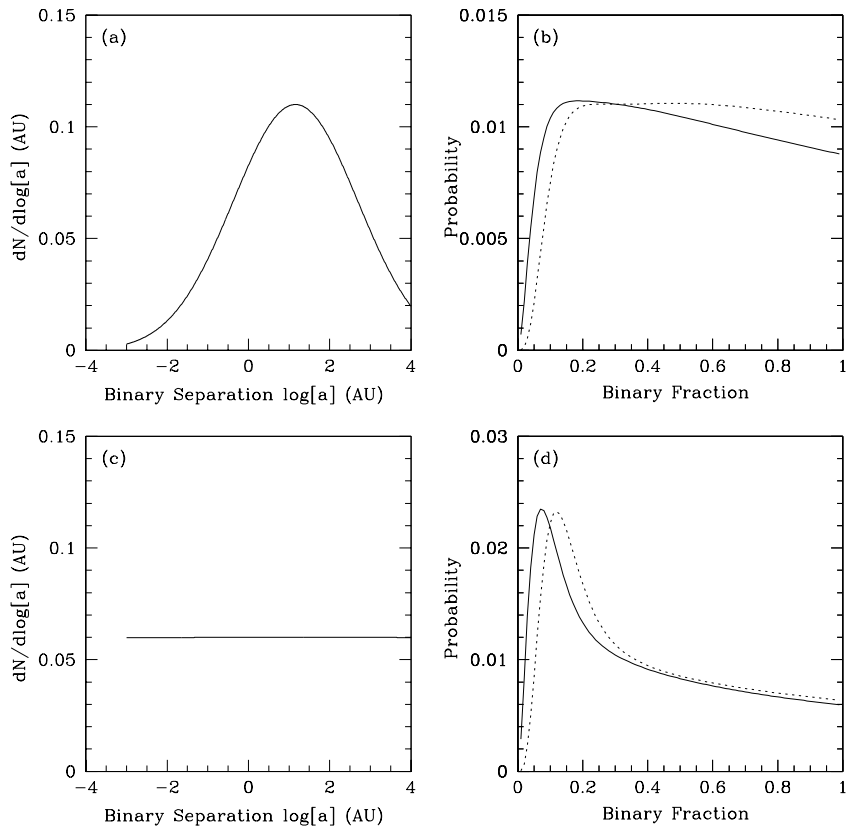}

\caption{On the left we show the assumed distribution of binary
  separations. The top plot has a peak at 15\,au and a Gaussian sigma
  of 1.53 in logarithmic units. They have been normalised to show the
  approximate binary fractions seen in sample of nearby M-dwarfs (see
  Fischer \& Marcy 1992, figure~2), with a total binary fraction of
  0.42. The bottom panel shows a flat distribution with a similar normalisation.
  On the right we show the
  corresponding probability that these distributions produce the
  observed numbers of binary systems. The solid curves show the results
  for sample A (2 binaries from 57), and the dashed curves show the
  results for sample B (4 binaries from 64).}
\label{binplots}
\end{figure*}

The data we have obtained offer us the opportunity to investigate the
fraction of VLMS and BDs in the $\sigma$~Ori cluster which are in short
period binary systems. A test can be formulated to see whether the RVs
of cluster members are (a) constant, if measured more than once, and
(b) consistent with the mean cluster RV. The fraction of short period
binaries identified in this way can then be compared with models of the
binary fraction as a function of separation, $a$, and mass ratio, $q$,
that also take into account the RV sampling and uncertainties.

A difficulty here is deciding which candidate VLMS and BDs to include as
genuine members. Clearly we cannot use measured RV as a criterion!  We
make two choices, one more restrictive than the other. The first sample
is all those VLMS/BD candidates which show clear evidence for Li in
their spectra -- a total of 57 objects, which we will call sample~A. 
The second sample, includes the
first as a subset, but also includes all those objects with an EW[Na]
which would seem to unambiguously classify them as low-gravity objects
-- a total of 64 objects, which we will label as sample~B. 
Since these bracket a range of possibilities
for the binary fraction, we perform calculations using both.

We applied a $\chi^2$ test to the radial velocities of the candidate
cluster members in each sample in order to identify binary systems. In
effect, we were testing the goodness-of-fit of the mean cluster RV to
the observed RV(s), adding the uncertainty in the cluster mean in
quadrature to each error bar. The $\chi^{2}$ threshold was set so that
there was a probability of $<10^{-3}$ of a constant RV cluster member
failing the test.  Two objects from sample~A (targets 46 and 74) were
identified as binary and a further 2 objects (targets 52, and 72)
in sample~B.  Then, assuming circular orbits and primary masses taken
from Fig.~\ref{cmd}, we used the mean times of observation and the RV
error bars for each target to calculate the probability that a binary
RV variation would have been detected at this level, both as a function
of $a$ and $q$ and averaged over all inclination angles.  Summing these
probabilities over all targets and over particular distributions of $q$
yields a ``detection efficiency'' as a function of binary separation
for that particular sample. We can then calculate the probability that
a given binary fraction, $q$ and $a$ distribution can result in the
number of binaries observed (see Maxted et al. 2001).

As an example, Fig.~\ref{subfig1} shows how efficiently (on average) we
can detect binary systems for a given separation for samples~A and~B. In the
absence of any definitive information on the distribution of $q$ in
VLMS/BDs we assume a flat mass ratio distribution for $0.2<q\leq 1$
which is reasonably consistent with what is observed for M-dwarfs
(Fischer \& Marcy 1992). If the $q$ distribution were peaked
towards $q=1$ then binaries become more detectable and the required
binary fraction to explain a set of observations becomes lower (and
vice-versa).  It is clear from this plot that our survey is not
sensitive to systems with separations of $\log a\ga 0$ (in au).  Conversely our
detection efficiency for closely separated binaries is very high.  

Given the sparse nature of our RV dataset we limit ourselves to an
attempt to answer one simple question. Is the number of candidate
binary systems we have observed consistent with the overall binary
fraction ($42\pm9$ per cent) and Gaussian $\log a$ distribution proposed
by Fischer \& Marcy (1992) for M-dwarfs within 20\,pc?  In one sense we
already know that for BDs the answer to this is no. There are too
few field and cluster BD-BD binary systems with $a>20$\,au compared
with M-dwarf binaries in the same separation range.
(e.g. Mart{\' \i}n et al. 2003; Gizis et al. 2003). However, we are probing
the small separation end of this distribution which is a separate
problem. We begin by using Fig.~\ref{subfig1} to calculate the
probability that a given binary fraction would have resulted in the
number of observed binary detections, as a function of $\log a$.
We then convolve this with an assumed separation distribution
that is a Gaussian with a peak at $\log a=1.16$ and a
sigma of 1.53 (in au)\footnote{Fischer \& Marcy favoured a similar {\it period}
distribution to that of G-dwarfs found by Duquennoy \& Mayor
(1991). Here we have done the same which implies that the peak position of the $a$
distribution should be reduced by a factor of two, from 30\,au to
15\,au, to account for the lower system mass among VLMS/BDs}.  The
normalisation of the function is effectively set by the number of
binaries we have observed. A plot of probability versus binary fraction
can then be produced by integrating over any given range of $\log a$,
although as the detection efficiency becomes negligible
for $\log a>0$ (see Fig.~\ref{subfig1}) we confine ourselves to $-3 < \log a < 0$.

In the case of sample~A and assuming that the Gaussian $\log a$
distribution shown in Fig.~\ref{binplots}a is appropriate, we find that
the most likely binary fraction (between $-3 < \log a < 0$ only) is
0.19, with a lower limit of $>0.13$ at a 90 per cent confidence level
(see Fig.~\ref{binplots}b).  This is inconsistent with the
estimate of about 5-7 per cent binarity for this range of $\log a$
(Marcy \& Benitz 1989; Fischer \& Marcy 1992) in nearby M-dwarfs.  For
sample~B where there is an even higher {\em observed} binary fraction
the probability curve peaks at a binary fraction of 0.47 and with a
lower limit of $>0.17$ at a 90 per cent confidence
level.

These results imply that if VLMS/BDs do have a separation distribution
that decreases towards smaller separations in a similar way to field
dwarfs, then we have observed too many to be consistent with the small
binary fraction found for field M-dwarfs at small separations. This in
turn means that either the distribution of separations is quite
different in the VLMS/BDs and/or the binary fraction at small
separations is very high.  To test the sensitivity to the assumed
distribution of $\log a$, we have performed another calculation
assuming a flat distribution (see Fig.~\ref{binplots}c). In this case
sample A returns a most probable binary fraction ($-3< \log a <0$) of
only 0.07, with a lower limit of 0.07 at 90 per cent confidence. So,
depending on the detail of the separation distribution it seems that
fraction of VLMS/BDs in short period binary systems could be quite
small (if the $\log a$ distribution is much flatter than found for
higher mass stars) or could certainly be as high as the 0.5 suggested
by Pinfield et al. (2003) for VLMS/BDs in the Pleiades. Either way the
binary properties for short-period VLMS/BDs would be quite
different from those of M-dwarfs, as indeed they are at larger
separations.

There are several caveats to this interesting conclusion. (i) It is
based on very small number statistics, which although statistically
taken account of in our calculations, render the result vulnerable to 
contamination of the binary sample. (ii) Greater consistency with
the M-dwarf distribution would be achieved if some of the candidate
binary members were not binaries. This might be the case if our RV
errors were underestimated, but in that case we would be less sensitive
to binaries and the deduced binary fractions would remain similar. (iii) The
binary fraction would be smaller if some of the binary candidates turn
out not members of the cluster. This certainly applies to sample B
which contains two ``grey area'' binary candidates for which we are not certain
of cluster membership. Indeed the large implied binary fraction could
be viewed as evidence that this is the case. However our main conclusion
holds even if we consider just the sample in which Li was detected.
(iv) If the $q$ distribution for VLMS/BDs was sharply peaked
towards $q=1$ this would make it easier to detect binaries and thus our
efficiency would be greater and the required binary fraction lower.

Our results seem reasonably consistent with the survey of 26 field
VLMS/BDs conducted by Guenther \& Wuchterl (2003). They used 2-3 RV
measurements with precision 0.1-1\,km\,s$^{-1}$ spread over 1-2 months.
They find 3 binaries in this sample and hence deduce a binary fraction
of at least 12 per cent for periods of $<100$~days (roughly equivalent
to 0.2\,au). If we perform our calculation for $-3 < \log a < -0.7$ for
the case of sample A and the Fischer \& Marcy (1992) distribution, we
obtain a most probable binary fraction of 11 per cent. 
Both of these numbers should be
compared with the $\sim 3$ per cent binarity implied by the Fischer \&
Marcy analysis over a similar separation range. 
To make further and more decisive progress would require a multi-epoch
RV survey of a large sample of VLMS/BDs in the $\sigma$~Ori
cluster and elsewhere. Observations spread over several months and with
a factor of a few better precision would give sensitivity to
separations as large as the minimum observable with adaptive optics
systems for nearby BDs {\em and} would allow some estimation of 
the $\log a$ distribution.

\subsection{Disc Fractions}
\label{discs}

The frequency and lifetimes of discs around young, VLMS/BDs may be an
important diagnostic tool in identifying the dominant (sub)stellar
formation scenario. A shorter disc lifetime for lower mass objects as a
result of dynamical ejection of stellar embryos from their initial gas
reservoirs is qualitatively supported by numerical models (Bate
et~al. 2003).  Haisch et~al. (2001) determined disc fractions (using
$K-L$ near infrared excess) for low-mass stars (0.2-1\,$M_{\odot}$) in
several clusters, demonstrating a rapid decrease in disc frequency with
age -- initial frequencies may be as high as $\sim 80$ per cent, with
approximately half of the discs being lost within $\la3$\,Myr, and an
overall lifetime of $\sim6$\,Myr. Thus, at an age of $3-7$\,Myr some
dissipation of circumstellar discs is certainly expected within the
$\sigma$~Ori cluster. Using similar diagnostics, Oliveira, Jeffries \& van
Loon (2004) found that $\sim 50$ per cent of low-mass stars around
$\sigma$~Ori possessed circumstellar material, broadly agreeing with
this scenario. Liu et al. (2003) and Jayawardhana et al. (2003b)
present $K-L$ measurements for BDs in a number of clusters which
support a similar timescale for disc dissipation in lower mass objects.

Five or possibly six of our very low-mass $\sigma$~Ori cluster members show 
evidence for the presence of accretion on the 
basis of broad H$\alpha$ emission. Three of these objects are probably
substellar according to Fig.~\ref{cmd}. The deduced accretion frequency of
$10\pm5$ per cent is much lower than the frequency of discs found
among higher mass objects by Oliveira et al. (2004). However, different
disc indicators cannot be considered equivalent. Natta et al. (2004)
show that a broadened H$\alpha$ profile with a full width at 10 per
cent of maximum of $>270$\,km\,s$^{-1}$ is sensitive to accretion rates
$>10^{-11}$\,$M_{\odot}$\,yr$^{-1}$ (for VLMS/BDs), whereas 
a $K-L$ excess  can be seen even from a passive (reprocessed light only) disc with
$M>10^{-6}$\,$M_{\odot}$ (Wood et al. 2002; Walker et
al. 2004). For example, 23 of the brighter objects in our sample (all with $I<16.75$)
have $K-L$ measurements in Oliveira et al. (2004) or Oliveira et
al. (in preparation). Of these, 6 to 9 (depending on the exact criteria) 
show a $K-L$ excess, but only targets 10, 13 and 37 have been identified as accretors. 
Jayawardhana et al. (2003b) also find that 2 out of 6
BDs in $\sigma$~Ori have $K-L$ excesses, but we have found no
evidence for accretion in either of these (S Ori 12 and S Ori 40).

A more sensible comparison would be with the frequency of strong,
accretion-related H$\alpha$ emission in stars of higher mass in the
$\sigma$~Ori cluster and VLMS/BDs in younger and older clusters. Among
younger clusters, Jayawardhana et al. (2003a) find the fraction of
accreting substellar objects (based on the width of the H$\alpha$ line)
to be about 50 per cent for ages $\leq 2\,$Myr (in IC\,348). Muzerolle
et al. (2003) measured a small sample of eight $\sigma$~Ori candidates,
finding no accretors. S Ori 12, 29 and 40 are in common with
this study and we also find no evidence for accretion in these
objects.  Neither do they find any VLMS/BD accretors in the older
($\sim 8$\,Myr) Sco-Cen association. Our measurements from a much larger
sample clarify this trend and pin down the accretion timescale for
most VLMS/BDs to be less than the age of the $\sigma$~Ori cluster.

The H$\alpha$ results for VLMS/BDs in the $\sigma$~Ori cluster are
similar to those for higher mass stars. Wolk (1996) states
that 12 per cent (6 out of 49) among a sample of $\sigma$~Ori candidates with
spectral types K0-M0 are classical T-Tauri stars with accretion discs,
using EW$[$H$\alpha]>10$\AA\ as his criterion. Using the same H$\alpha$
criterion, Zapatero-Osorio et al. (2002) find one star from 11 shows
accretion in the same spectral type range.  Among younger clusters, the
preponderance of accretion-induced H$\alpha$ emission is much
higher. For example, 63 per cent (30/48) stars with $M>0.2\,M_{\odot}$
are accretors in the Taurus-Auriga (age 1-2\,Myr) sample of White \&
Ghez (2001).

Hence, it seems that, similarly to the $K-L$ disc indicator, the
fraction of objects with discs betrayed by their accretion signatures
declines at a rate that is roughly independent of mass, but that the
decline takes place more rapidly. Interpretation of this result
is complicated by the mass-dependence of the sensitivity of these
tests. Most low-mass classical T-Tauri stars accrete at rates that are
orders of magnitude higher than the threshold below which their
accretion would be undetectable. Accreting BDs on the other hand are
detected at rates right down to a smaller threshold. This suggests that
the transition between accreting and non-accreting status may happen
quite rapidly for low-mass stars and that the fraction of accreting BDs
may be an underestimate.  Either way it seems that accretion may become
undetectable even whilst a significant mass of disc material is still
present. If accretion at a rate $\sim 10^{-11}$\,$M_{\odot}$\,yr$^{-1}$
is present in BDs at an age of $\sim 5$\,Myr, it implies disc masses of
order $5\times10^{-5}\,M_{\odot}$. The recent models of flared discs
around BDs by Walker et al. (2004), do indeed suggest that near-IR
excesses should be present for these and even less massive discs.

The presence of accretion discs in a significant fraction of VLMS/BDs
at an age of a few Myr and in similar numbers to those seen in higher
mass stars at the same age, {\em does not} support the idea of VLMS/BDs
forming by a dynamical ejection process in which their discs are
truncated. Instead it argues for a common formation mechanism for
low-mass stars, VLMS and BDs. However, whether such observations can
definitively rule out the ejection model must await detailed
predictions of the initial BD disc properties and masses from such a
scenario (e.g. see the discussion in Liu et al. 2003 and Natta et al. 2004).

\section{Conclusions}

We have used intermediate resolution spectroscopy to refine a set of
candidate VLMS and BD members of the $\sigma$~Ori cluster (age $\sim
3-7$\,Myr) that were chosen
on the basis of their $RI$ photometry. We have measured radial
velocities, the strengths of absorption lines due to neutral lithium
and sodium and searched for signs of accretion activity. 

We find that the majority of the photometrically selected candidates
{\em are} genuine cluster members on the basis that they exhibit an
Li~{\sc i}~6708\AA\ absorption feature, have radial velocities
consistent with cluster membership and have a Na~{\sc i}~8183/8195\AA\
absorption doublet that is weaker than main sequence M-dwarfs of a
similar spectral type. The cluster members we have identified have
model- and distance-dependent masses of $0.03<M<0.35\,M_{\odot}$. Approximately
23 of our targets are likely to have substellar masses when
measured in this way.

We have found another class of candidate where the evidence is
ambiguous. These objects either have upper limits to their Li line EWs
that may be consistent with cluster membership, have radial
velocities that indicate they are non-members or cluster
binaries or have no Li but radial velocities consistent with cluster membership.
Several of these objects have weak Na~{\sc i} absorption that
indicates a youthful status. These objects, together with the apparent
spread in the strength of the Li~{\sc i}~6708\AA\ absorption
EW among cluster members, lead us to conclude
that there may be deficiencies in our understanding of the formation of
this feature in cool, low gravity objects. It may be unsafe to rule out
the membership of a VLMS or BD from a young cluster or association solely
on the basis of a weak or absent Li feature.

Despite these problems we find that overall, photometric selection
alone is reasonably effective in isolating members of the $\sigma$~Ori
cluster. We estimate contamination rates of 15-25 per cent among the
higher mass stars of our sample and $<20$ percent at the lower mass
end. Contamination at this level will not significantly change the form
of the cluster mass function that has been deduced from photometrically
selected samples. However, we do show that selection in the $I$ versus
$I-J$ diagram is potentially more discriminating than the $I$ versus
$R-I$ diagram.

We have identified 4 candidate binary cluster members on the basis of a
varying RV or an RV discrepant from the cluster mean combined with
other indications of membership. Two of these candidates show signs of
Li in their spectra. A careful consideration of the RV errors and
sampling in our data enables us to determine how efficiently we can
detect binary systems as a function of primary mass, mass ratio and
separation. Combining this calculation with the number of binaries we
have observed and plausible distributions of binary mass ratio and
separation, we determine that we have found too many binary systems to
be accounted for by binary properties similar to those in higher mass
stars. Instead, the short-period binary frequency could be much larger in
our VLMS/BDs (perhaps as high 0.5 for $-3 < \log a < 0$) {\em or} the
binary frequency is lower but the distribution with $\log a$ must be
flatter than seen in higher mass field M-dwarfs. Caveats to this conclusion
must be considered. The results are crucially dependent on the correct 
identification of a small number of binary members
of the cluster. However, these conclusions still apply even if we
restrict ourselves to considering cluster members where the presence of
Li has been positively identified. Further work will be needed to hone
this result -- a multi-epoch RV survey of a large sample of
$\sigma$~Ori VLMS/BDs with a better RV precision has the capability to
ascertain the binary frequency and separation distribution at
separations smaller than accessible by adaptive optics observations of
nearby BDs.

A search for broadened H$\alpha$ emission features associated with a
circum(sub)stellar accretion disc has revealed 5 or 6 accreting objects, 3 of
which are found among the substellar candidates. The fraction of
VLMS/BD accretors ($10\pm5$ per cent) is similar to that found among
higher mass stars in the cluster using similar diagnostics, but much
lower than the fraction found amongst low-mass stars, VLMS and BDs in
younger clusters. The timescale for accretion rates to drop to
$<10^{-11}$\,$M_{\odot}$\,yr$^{-1}$ is thus constrained to be less than
the age of the $\sigma$~Ori cluster for most VLMS/BDs. The similar
timescale for a decline in accretion rates for both VLMS/BDs and higher
mass stars does not support the idea that VLMS/BDs have truncated,
lower mass discs as a result of dynamical ejection. However, in the
absence of detailed predictions of disc masses and properties for this
scenario, it cannot be completely ruled out. Instead the
observations are consistent with the notion that low-mass stars and BDs
form a continuum of outcomes from a single formation process in which
the presence of circumstellar material is a natural consequence.

\section*{Acknowledgements}
This research
has made use of data obtained from the Leicester Database and Archive
Service at the Department of Physics and Astronomy, Leicester
University, UK. The Digitized Sky Survey was produced at the Space
Telescope Science Institute under U.S. Government grant NAG W-2166. The
images of these surveys are based on photographic data obtained using
the Oschin Schmidt Telescope on Palomar Mountain and the UK Schmidt
Telescope. The plates were processed into the present compressed
digital form with the permission of these institutions. 
We thank the director and staff of the the William Herschel and Isaac
Newton telescopes, which are operated on the island of La Palma by the
Isaac Newton Group in the Spanish Observatorio del Roque de los
Muchachos of the Instituto de Astrofisica de Canarias.  
MJK acknowledges the support of a PPARC research studentship. JMO 
acknowledges the financial support of PPARC. We are grateful to Mike
Irwin and the Cambridge Astronomical Survey Unit for advice and 
calibration frames for the photometric data.


\bsp 

\include{bigtable}

\label{lastpage}

\end{document}

%% file: bigtable.tex
\begin{table*}
\tiny
\caption{Positions, $I$ magnitudes, $I-J$ and $R-I$ colours, heliocentric
radial velocities (nights 1 and 2) and measured equivalent widths for
our targets. Photometric uncertainties in $I$ and $R-I$ are about
$\pm0.04$ and $\pm 0.05$ respectively (see section 2). $J$ magnitudes
are from 2MASS. Alternate names from B\'{e}jar et al. (2001) are
given. The final three columns indicate our classification of
membership according to whether the EW[Na] indicates a low gravity (Y)
or is ambiguous (?); whether the RV is consistent with membership and
finally whether an object is classed as a member (Y), non-member (N) or
is in a grey-area (?). Objects with broad H$\alpha$ emission or which
are cluster binary candidates are labelled with an a (for accretor) or b.}
\label{rvmembers}
\begin{tabular}{@{}lccccccccclll}
\hline
ID&RA&DEC&$I$&$I-J$&$R-I$&RV$_1$&RV$_2$&\lith&\sod&Low &RV&$\sigma$Ori\\
&(J2000)&(J2000)&&&&(km/s)&(km/s)&(\AA)&(\AA)&$g$& &Member?\\
\hline
 & & & & & & & & & & & \\ 
 1                                &5 38 43.75&-2 52  42.8&14.81 &1.79 &1.78 & \ldots  & 33$\pm$4&0.56$\pm$0.08&2.05$\pm$0.18&Y&Y&Y\\ 
 2                                &5 39 40.97&-2 16  24.3&14.84 &1.97 &1.83 & \ldots  & 30$\pm$3&0.49$\pm$0.03&1.50$\pm$0.12&Y&Y&Y\\ 
 3                                &5 38 17.46&-2 09  23.6&14.94 &1.67 &1.77 & \ldots  & 25$\pm$3&0.71$\pm$0.05&2.31$\pm$0.15&Y&Y&Y\\ 
 4                                &5 39 05.24&-2 33  00.5&14.97 &1.58 &1.60 & \ldots  & 36$\pm$3&0.58$\pm$0.06&1.80$\pm$0.12&Y&Y&Y\\ 
 5                                &5 38 44.49&-2 40  30.4&14.98 &1.62 &1.52 & \ldots  & 29$\pm$4&0.50$\pm$0.04&2.41$\pm$0.17&Y&Y&Y\\ 
 & & & & & & & & & & & \\ 
 6                                &5 40 01.02&-2 19  59.8&15.02 &1.92 &1.90 & \ldots  & 29$\pm$4&0.56$\pm$0.08&1.61$\pm$0.15&Y&Y&Y\\ 
 7                                &5 38 10.34&-2 13  59.6&15.03 &1.56 &1.52 & \ldots  & 71$\pm$3&   $<$0.07   &2.88$\pm$0.17&?&N&N\\ 
 8                                &5 38 50.76&-2 36  26.6&15.06 &1.95 &2.00 & \ldots  & 29$\pm$4&0.22$\pm$0.05&1.52$\pm$0.12&Y&Y&Y\\ 
9                                &5 39 01.15&-2 36  38.7&15.11 &1.59 &1.56 & \ldots  & 30$\pm$3&0.62$\pm$0.03&1.95$\pm$0.12&Y&Y&Y\\
10                                &5 40 09.34&-2 25  6.75&15.11 &1.96 &1.57 & \ldots  & 33$\pm$3&0.37$\pm$0.07&1.69$\pm$0.15&Y&Y&Y a\\
 & & & & & & & & & & & \\ 
11                                &5 38 16.09&-2 38  4.87&15.17 &1.59 &1.62 & \ldots  & 34$\pm$2&0.52$\pm$0.04&1.74$\pm$0.12&Y&Y&Y\\
12                                &5 39 13.48&-2 23  51.8&15.18 &1.23 &1.41 & \ldots  & 43$\pm$3&   $<$0.07   &2.17$\pm$0.14&?&N&N\\
13                                &5 38 23.59&-2 20  47.6&15.24 &1.83 &1.56 & \ldots  & 30$\pm$2&0.48$\pm$0.04&2.23$\pm$0.15&Y&Y&Y a\\
14                                &5 37 52.09&-2 56  55.1&15.24 &1.85 &1.77 & 25$\pm$6& 25$\pm$3&0.64$\pm$0.04&2.42$\pm$0.17&Y&Y&Y\\
15                                &5 38 23.32&-2 44  14.2&15.27 &1.81 &1.58 & \ldots  & 23$\pm$3&0.61$\pm$0.04&2.40$\pm$0.15&Y&Y&Y\\
 & & & & & & & & & & & \\ 
16                                &5 36 46.92&-2 33  28.3&15.28 &1.73 &1.71 & \ldots  & 28$\pm$3&0.48$\pm$0.08&2.59$\pm$0.47&Y&Y&Y\\
17 (SOri6)                        &5 38 47.65&-2 30  37.4&15.33 &1.88 &1.92 & 39$\pm$5& \ldots  &0.55$\pm$0.04&2.08$\pm$0.17&Y&Y&Y\\
18                                &5 37 58.39&-2 41  26.1&15.36 &2.07 &1.84 & 30$\pm$2& 37$\pm$2&0.61$\pm$0.04&2.04$\pm$0.17&Y&Y&Y\\
19                                &5 39 50.56&-2 34  13.7&15.39 &1.71 &1.69 & 34$\pm$2& 29$\pm$4&0.56$\pm$0.04&2.15$\pm$0.14&Y&Y&Y\\
20 (SOri3)                        &5 39 20.97&-2 30  33.5&15.40 &2.11 &1.98 & 37$\pm$3& \ldots  &0.35$\pm$0.05&1.86$\pm$0.15&Y&Y&Y\\
 & & & & & & & & & & & \\ 
21                                &5 38 54.93&-2 28  58.3&15.44 &1.64 &1.74 & \ldots  & 30$\pm$2&0.57$\pm$0.05&2.28$\pm$0.20&Y&Y&Y\\
22                                &5 38 40.07&-2 50  37.1&15.44 &1.77 &1.75 & 36$\pm$2& \ldots  &0.55$\pm$0.05&1.95$\pm$0.17&Y&Y&Y\\
23                                &5 37 54.52&-2 58  26.4&15.51 &2.20 &1.93 & 36$\pm$4& \ldots  &0.46$\pm$0.07&1.37$\pm$0.15&Y&Y&Y\\
24                                &5 39 42.99&-2 13  33.3&15.53 &1.63 &1.73 & 25$\pm$3& \ldots  &0.53$\pm$0.03&2.13$\pm$0.15&Y&Y&Y\\
25                                &5 37 50.32&-2 12  24.8&15.56 &1.66 &1.62 & 34$\pm$3& 43$\pm$4&   $<$0.19   &2.39$\pm$0.21&Y&Y&?\\
 & & & & & & & & & & & \\ 
26                                &5 39 15.06&-2 18  44.4&15.59 &1.38 &1.61 &120$\pm$5& \ldots  &   $<$0.07   &2.80$\pm$0.20&?&N&N\\
27                                &5 37 56.14&-2 09  26.7&15.68 &1.78 &1.78 & 41$\pm$3& \ldots  &   $<$0.55   &1.48$\pm$0.45&Y&Y&? \\
28                                &5 37 49.65&-2 49  02.9&15.70 &1.42 &1.66 & \ldots  & -2$\pm$3&   $<$0.06   &3.12$\pm$0.20&?&N&N\\
29                                &5 40 34.60&-2 33  13.8&15.72 &1.55 &1.50 & \ldots  & 30$\pm$2&   $<$0.08   &2.84$\pm$0.18&?&Y&?\\
30                                &5 39 05.82&-2 26  15.4&15.76 &1.42 &1.53 &-25$\pm$3& \ldots  &   $<$0.07   &3.16$\pm$0.21&?&N&N\\
 & & & & & & & & & & & \\ 
31                                &5 38 50.61&-2 42  42.9&15.90 &2.06 &1.86 & 31$\pm$3& \ldots  &0.40$\pm$0.04&2.10$\pm$0.15&Y&Y&Y\\
32                                &5 38 17.00&-2 14  46.5&15.97 &1.73 &1.94 & 25$\pm$4& \ldots  &0.24$\pm$0.11&2.64$\pm$0.20&Y&Y&Y\\
33 (J053909.9-022814)             &5 39 10.01&-2 28  11.6&16.05 &1.45 &1.63 & \ldots  & 32$\pm$3&   $<$0.06   &2.85$\pm$0.18&?&Y&?\\
34                                &5 38 18.28&-2 43  35.4&16.13 &1.45 &1.33 & \ldots  & 16$\pm$9&   $<$0.17   &3.12$\pm$0.38&?&Y&?\\
35                                &5 38 48.19&-2 44  00.8&16.15 &2.08 &1.86 & \ldots  & 47$\pm$7&0.53$\pm$0.10&1.68$\pm$0.21&Y&Y&Y\\
 & & & & & & & & & & & \\ 
36                                &5 38 33.89&-2 45  07.9&16.17 &1.92 &1.78 &24$\pm$5 & \ldots  &0.38$\pm$0.07&2.17$\pm$0.15&Y&Y&Y\\ 
37 (J053849.2-022358)             &5 38 49.28&-2 23  57.6&16.27 &1.91 &1.74 & 42$\pm$5& \ldots  &0.49$\pm$0.05&1.20$\pm$0.15&Y&Y&Y a\\
38                                &5 37 28.64&-2 46  46.9&16.28 &1.47 &1.50 & -6$\pm$3& \ldots  &   $<$0.09   &3.31$\pm$0.25&?&N&N\\
39 (SOri12)                       &5 37 57.45&-2 38  44.4&16.28 &2.05 &1.87 & 37$\pm$2& \ldots  &0.54$\pm$0.06&2.23$\pm$0.20&Y&Y&Y\\
40 (SOri10)                       &5 39 44.51&-2 24  43.2&16.30 &2.21 &1.72 & \ldots  & 35$\pm$6&   $<$0.23   &1.94$\pm$0.15&Y&Y&?\\
 & & & & & & & & & & & \\ 
41 (SOri15)                       &5 38 48.09&-2 28  53.6&16.39 &1.92 &1.97 & \ldots  & 39$\pm$5&0.46$\pm$0.14&2.03$\pm$0.18&Y&Y&Y\\
42 (J053911.4-023333)             &5 39 11.40&-2 33  32.7&16.42 &1.97 &1.89 & 35$\pm$5& \ldots  &0.33$\pm$0.06&1.88$\pm$0.30&Y&Y&Y\\
43                                &5 38 10.12&-2 54  50.6&16.45 &2.06 &1.93 & \ldots  & 32$\pm$7&0.25$\pm$0.05&1.21$\pm$0.28&Y&Y&Y\\
44                                &5 38 38.59&-2 41  55.9&16.47 &1.91 &1.80 & \ldots  & 34$\pm$3&0.55$\pm$0.06&2.54$\pm$0.18&Y&Y&Y\\
45                                &5 40 17.05&-2 26  48.9&16.51 &1.52 &1.60 & 32$\pm$3& 38$\pm$4&   $<$0.09   &3.00$\pm$0.23&?&Y&?\\
 & & & & & & & & & & & \\ 
46                                &5 40 00.10&-2 51  59.4&16.57 &2.43 &2.05 & \ldots  & 17$\pm$3&0.25$\pm$0.09&2.18$\pm$0.21&Y&N&Y b\\
47 (SOri18)                       &5 38 25.68&-2 31  21.7&16.61 &1.94 &1.91 & 34$\pm$4& 32$\pm$3&0.30$\pm$0.10&2.15$\pm$0.18&Y&Y&Y\\
48                                &5 38 13.30&-2 51  33.0&16.62 &2.05 &1.95 & 28$\pm$4& \ldots  &0.68$\pm$0.13&1.87$\pm$0.17&Y&Y&Y\\
49                                &5 39 37.59&-2 44  30.3&16.75 &2.37 &1.94 & 40$\pm$5& \ldots  &0.71$\pm$0.06&1.94$\pm$0.37&Y&Y&Y\\
50                                &5 38 51.00&-2 49  13.9&16.79 &1.75 &1.79 & 28$\pm$5& \ldots  &0.33$\pm$0.07&2.61$\pm$0.23&Y&Y&Y\\
 & & & & & & & & & & & \\ 
51 (SOri22)                       &5 38 35.36&-2 25  22.2&16.88 &2.23 &2.04 & \ldots  & 32$\pm$3&1.00$\pm$0.18&2.01$\pm$0.20&Y&Y&Y\\
52 (SOri20)                       &5 39 07.58&-2 29  5.61&16.96 &2.00 &2.00 & 70$\pm$3& \ldots  &   $<$0.13   &3.64$\pm$0.28&Y&N&? b\\
53                                &5 38 46.58&-2 19  40.5&16.97 &1.66 &1.75 & 22$\pm$3& \ldots  &   $<$0.08   &3.73$\pm$0.29&?&Y&?\\
54                                &5 40 32.48&-2 40  59.8&16.97 &2.27 &2.07 & 31$\pm$4& 19$\pm$4&0.49$\pm$0.12&3.05$\pm$0.26&Y&Y&Y\\
55 (SOri29)                       &5 38 29.62&-2 25  14.2&17.02 &2.18 &2.00 & 28$\pm$4& \ldots  &0.32$\pm$0.14&1.82$\pm$0.20&Y&Y&Y\\
 & & & & & & & & & & & \\ 
56                                &5 38 47.14&-2 57  55.7&17.05 &2.54 &2.14 & 33$\pm$6& 42$\pm$4&0.53$\pm$0.10&1.93$\pm$0.38&Y&Y&Y a\\
57                                &5 37 27.61&-2 57  10.0&17.06 &1.95 &1.80 & 42$\pm$7& 46$\pm$8&0.48$\pm$0.11&3.14$\pm$0.45&Y&Y&Y\\
58                                &5 38 26.24&-2 40  41.3&17.09 &2.19 &2.00 & 36$\pm$4& \ldots  &0.51$\pm$0.08&1.86$\pm$0.24&Y&Y&Y\\
59                                &5 39 43.39&-2 53  23.0&17.15 &1.86 &1.72 & 24$\pm$5& 31$\pm$4&   $<$0.08   &3.87$\pm$0.34&?&Y&?\\
60 (SOri27)                       &5 38 17.41&-2 40  24.2&17.20 &2.37 &2.00 & \ldots  & 47$\pm$5&0.31$\pm$0.06&2.86$\pm$0.20&Y&Y&Y \\
 & & & & & & & & & & & \\ 
61 (SOri21)                       &5 39 34.33&-2 38  46.8&17.25 &2.49 &2.02 & 26$\pm$4& 29$\pm$6&0.63$\pm$0.14&2.46$\pm$0.23&Y&Y&Y\\
62                                &5 37 52.06&-2 36  04.7&17.30 &2.16 &1.91 & \ldots  & 25$\pm$4&0.81$\pm$0.19&3.45$\pm$0.45&Y&Y&Y\\
63                                &5 40 13.95&-2 31  27.3&17.34 &2.77 &2.19 & 30$\pm$4& 30$\pm$5&0.47$\pm$0.08&2.41$\pm$0.23&Y&Y&Y a\\
64 (SOri28)                       &5 39 23.19&-2 46  55.5&17.36 &2.03 &2.05 & 29$\pm$3& 24$\pm$4&0.66$\pm$0.09&2.46$\pm$0.20&Y&Y&Y\\
65                                &5 38 39.76&-2 32  20.3&17.58 &2.69 &1.89 & 31$\pm$7& 36$\pm$5&0.14$\pm$0.05&1.79$\pm$0.18&Y&Y&Y\\
 & & & & & & & & & & & \\ 
66 (J053821.3-023336)             &5 38 21.38&-2 33  36.3&17.59 &2.24 &2.04 & \ldots  & 45$\pm$6&0.34$\pm$0.06&1.87$\pm$0.25&Y&Y&Y\\
67 (SOri32)                       &5 39 43.57&-2 47  31.7&17.64 &2.31 &2.11 & 33$\pm$5& 23$\pm$7&0.46$\pm$0.09&2.49$\pm$0.23&Y&Y&Y\\
68 (SOri39)                       &5 38 32.44&-2 29  57.2&17.67 &2.23 &2.09 & \ldots  & 19$\pm$9&0.45$\pm$0.10&3.68$\pm$0.31&Y&Y&Y\\
69                                &5 38 37.88&-2 20  39.8&17.81 &2.21 &2.14 & \ldots  & 26$\pm$3&0.49$\pm$0.14&3.58$\pm$0.28&Y&Y&Y\\
70                                &5 38 12.41&-2 19  38.8&17.81 &2.33 &2.16 & \ldots  &  7$\pm$8&0.38$\pm$0.12&2.17$\pm$0.48&Y&Y&Y\\
 & & & & & & & & & & & \\ 
71                                &5 37 42.47&-2 56  35.8&17.82 &1.92 &1.86 &  6$\pm$9& \ldots  &0.37$\pm$0.11&1.58$\pm$0.28&Y&Y&Y\\
72                                &5 37 39.66&-2 18  26.7&17.82 &2.38 &2.02 & -3$\pm$2& 32$\pm$3&   $<$1.16   &3.12$\pm$0.40&Y&N&? b\\
73                                &5 40 04.52&-2 36  42.0&17.86 &2.56 &2.07 & 42$\pm$6& 31$\pm$3&   $<$0.39   &2.84$\pm$0.31&Y&Y&? a\\
74 (SOri36)                       &5 39 26.85&-2 36  56.0&18.01 &2.55 &1.88 & 28$\pm$4& 15$\pm$4&0.45$\pm$0.15&2.80$\pm$0.26&Y&N&Y b\\
75 (SOri40)                       &5 37 36.48&-2 41  56.7&18.08 &2.61 &2.04 & \ldots  & 55$\pm$9&0.66$\pm$0.20&1.15$\pm$0.60&Y&Y&Y\\
 & & & & & & & & & & & \\ 
76                                &5 38 18.34&-2 35  38.5&18.15 &2.70 &2.16 & 31$\pm$9& 56$\pm$7&   $<$0.19   &2.09$\pm$0.32&Y&Y&?\\
\hline
\end{tabular}
\end{table*}